\documentclass[12pt]{JHEP3}
\usepackage{amsfonts,amssymb}
\usepackage{amsmath}

\usepackage{epsf}

\def\T11{{T}^{1,1}}
\def\bear{\begin{eqnarray}}
\def\eear{\end{eqnarray}}

\def\dim{\mathrm{dim}}
\newcommand{\vac}{{|0\rangle}}

\newcommand{\pa}{\partial}
\newcommand{\tr}{{\rm tr}}
\newcommand{\comment}[1]{}

\newcommand{\CO}{{\cal O}}
\newcommand{\pasl}{\pa\kern-.55em /}

\newcommand{\ksl}{k\kern-.55em /}

\newcommand{\ket}[1]{|#1\rangle}
\newcommand{\bra}[1]{\langle #1|}
\newcommand{\vev}[1]{\langle #1\rangle}

\DeclareFixedFont{\xiiss}{OT1}{cmss}{m}{n}{12}
\DeclareFixedFont{\ixss}{OT1}{cmss}{m}{n}{9}
\DeclareFixedFont{\cmrnine}{OT1}{cmr}{m}{n}{9}
\newcommand{\field}[1]{\mathbb{#1}}

\newcommand{\BC}{{\field C}}
\newcommand{\BR}{{\field R}}
\newcommand{\BZ}{{\field Z}}

\newcommand{\CCs}{\hbox{\ixss C\kern-.4emI}}
\newcommand{\ZZs}{\hbox{\ixss Z\kern-.4emZ}}

\newcommand{\CQA}{{\cal QA}}

\title{ Strings on conifolds from strong coupling dynamics, part I }
\author{David Berenstein$^{\dagger,1}$\\
$^\dagger$ Department of Physics, UCSB, Santa Barbara, CA 93106    }

\abstract{ A method to solve various aspects of the strong coupling expansion of the superconformal field theory duals of $AdS_5\times X$ geometries from first principles is proposed. The main 
idea is that at strong coupling the configurations that dominate the low energy dynamics of the 
field theory compactified on a three sphere are given by certain non-trivial semi-classical configurations in the moduli space of vacua. 
 We show that this approach is self-consistent and permits one to express most of the dynamics in terms of an effective ${\cal N}=4$ SYM dynamics. This has the advantage that some degrees of freedom that move the configurations away from moduli space can be treated perturbatively,  unifying the essential low energy dynamics of all of these theories. We show that with this formalism one can compute the energies of strings in the BMN limit in the Klebanov-Witten theory from field theory considerations, matching the functional form of  results found using AdS geometry. This paper also presents various other technical results for the semiclassical treatment of superconformal field theories.}

\begin{document}

\section{Introduction}

The AdS/CFT correspondence is one of the most interesting developments in modern theoretical physics \cite{Malda}. In the simplest cases it relates ordinary gauge field theories in four dimensions to quantum gravity theories in higher dimensions. The first non-trivial proposal for this correspondence is the case of ${\cal N}=4 $ SYM and type IIB string theory on $AdS_5\times S^5$. The ${\cal N}=4$ SYM theory is conformally invariant and has the maximal number of supersymmetries. This is the first example of a dual string theory in four dimensions for a large $N$ gauge theory, as suggested by 't Hooft \cite{'tH}

If one reduces the number of supersymmetries, 
one can find that certain examples of such a correspondence arise from pairs of a space $AdS_5\times X$, where $X$ is a Sasaki-Einstein manifold, and some particular superconformal field theory in four dimensions \cite{KW,MP}. If one looks a list of these conformal field theories ( a long list of AdS/CFT dual pairs can be found in \cite{HKW}), very little is known about them. They can be proven to exist by the arguments of Leigh and Strassler \cite{LS}. One finds this way that the dimensions of the ``fundamental" chiral fields are not equal to one. Thus these conformal field theories have large anomalous dimensions and can not be argued to be weakly coupled: they are not near a Gaussian fixed point where perturbation theory is valid. 

In order to test the AdS/CFT correspondence, one needs to compare the representation theory of the conformal group between the AdS side and the CFT side \cite{GKP,Wads}. Matching of the representation theory is considered a very strong test of the correspondence. In the CFT side this is accomplished by studying the spectrum of local operator and their scaling dimensions. This is the same problem as studying the spectrum of energies of the theory on $S^3\times \BR$ when one uses the operator-state correspondence of the theory. In this paper we will find it more convenient to take the second point of view rather than calculating the spectrum of operator dimensions.

 In the gravity side, the same representation theory can be understood by quantizing the gravity theory, plus quantizing various objects propagating in the geometry, in global coordinates. 

Holomorphy and symmetries permit one to calculate a few dimensions of composite operators exactly, in particular the dimension of the chiral ring operators. From the point of view of representation theory these are BPS operators. They saturate an inequality between the dimension and the R-charge, and they correspond to short representations of the superconformal group. For these operators, the R-charge 
is equal to the dimension of the operator (under a suitable choice of normalization of the R-charge). The chiral ring only depends on the superpotential and on identifying correctly the R-charge of the theory. Thus, one requires very little information to compute these objects.
In the dual gravitational side, the Sasaki-Einstein structure also predicts that $X$ is endowed with a $U(1)$ isometry. This is the dual realization of the $R$-charge symmetry that a superconformal field theory predicts. The BPS states that saturate the bound in gravity end up being given by particular supergravity excitations, and they can also be computed, even if one does not have a complete metric on $X$.

We should remember that a Sasaki-Einstein space $X$ can be understood as a base of a Calabi-Yau cone geometry $V$.  The dual theories to $AdS_5\times X$ are built by placing various D-branes at the tip of the cone of $V$ and taking the low energy limit of the open strings ending on the D-brane configurations. 

This is an indirect construction, because in general it is hard to understand how 
this can be done systematically, especially in cases where $V$ is not an orbifold (the case of orbifolds was solved in \cite{DM,KacS} by the method of images).
There are proposals for such theories when $V$ is a toric variety \cite{HKW}, and the data
that is obtained from these proposals is a quiver gauge theory with some matter content and
a superpotential. The rest of the theory is left implicit: it is understood that the rest should be determined automatically by the requirements of conformal invariance. 
One can also build such theories by taking orbifolds and further deforming the superpotential
by marginal and relevant operators. One then needs to flow to an IR non-trivial fixed point 
that would realize such a superconformal field theory. Again, this is one way to give a UV completion of the theory where some calculations can be performed (anomalies and holomorphy do not depend on many details of the UV theory), 
but the superconformal IR fixed point remains for the most part out of reach.

In general this is considered to be a very difficult problem: most of these IR fixed points theories are generically strongly coupled (in the sense that anomalous dimensions of fundamental chiral matter do not vanish). 

One would also like a way to understand how given a superconformal theory we would be able to determine the dual AdS geometry from first principles (if it exists). Under certain special
conditions on the superpotential, one can do this by using algebraic methods \cite{BL,Brev}.
This is expected to work only for $AdS\times X$ geometries without fluxes. When one turns on fluxes, there might be various possible geometric dual theories: a particularly interesting example is found in orbifolds with discrete torsion \cite{D,BLdt}, where one finds that certain marginal deformations of ${\cal N}=4 $ SYM are dual to $\BC^3/{\BZ_n\times \BZ_n}$ orbifolds. If one studies these deformations \cite{BJL,BJL2} one finds that the different possible geometric dual spaces that are dual to these field theories are related by T-dualities. More recently, the exact supergravity dual backgrounds have been constructed for the case of some particular small deformations \cite{LM}. For these more general cases very little is known about how to construct the dual geometries.

At first glance the collection of field theories that are used seem to have no relation to one
another, other than some rather general aspects of their structure. But the gauge groups and matter content can differ substantially between various theories. In light of this it is rather surprising that the dual AdS geometries are many times related to supergravity solutions of type IIB string theory. This is, there appears to be a universal dynamical principle for very different starting points when we are in the appropriate strong coupling regime.

So there is a question that begs to be asked: how does one unify very different looking field theories so that their dynamics is 
essentially the same at strong coupling? This is,  type IIB string theory, on a manifold with large radius of curvature where semiclassical gravity calculations are valid. This paper is geared towards answering this question for some restricted set of superconformal field theories.

In previous work, a new set of ideas to solve the strong coupling dynamics of ${\cal N}=4 $ SYM 
was proposed \cite{Droplet}. These ideas grew from a study of the chiral ring of the ${\cal N}=4 $ SYM theory. These generalized a previous classical treatment of half BPS states
\cite{Btoy} that were shown to be equivalent to an integer quantum hall system. Furthermore, the dual supergravity solutions of the half BPS states could also be described in terms of 
configurations of an incompressible fluid on a plane \cite{LLM}. 

 The basic premise of that work \cite{Droplet} is that at strong coupling
the ground state is dominated by configurations of spherically symmetric scalar field configurations (these are the only fields that are excited perturbatively to build chiral ring states).
 These are further constrained to be commuting matrices by the 
interactions (in particular, for BPS states, the matrices need to commute classsically to saturate the BPS inequalities). The ${\cal N}=4 $ SYM has six matrices, and via gauge transformations one
can diagonalize these matrices. 
The problem leads one to consider a  Schr\"odinger problem for a system of identical particles in six dimensions
with some interactions induced by the measure of the change of variables from general commuting matrices to diagonal matrices. The coordinates of the associated particles are the eigenvalues of the matrices themselves, and as the field theory has six scalar fields in the adjoint, this is how one gets six coordinates per particle (this is similar to \cite{BFSS}). After one solves an effective Schr\"odinger equation for these degrees of freedom, one can relate aspects of measurements of the wave functions to a Boltzmann gas of particles in six dimensions with logarithmic repulsions.
These techniques were designed to include the chiral ring information ab initio and they have been able to 
reproduce exactly the spectrum of BMN energies \cite{BMN} to all orders in the t'Hooft 
coupling \cite{BCV} (similar calculations were done in \cite{Joao}) . This field theory calculation also 
matched the calculation found in \cite{SZ} and the expected magnon dispersion relation in the strong coupling Bethe Ansatz \cite{BDS}. Furthermore, it
predicted various geometrical features of the giant magnons that 
were later found in the string theory dual classical configurations \cite{HM}. Also, recently the spectrum of some bound states of giant magnons \cite{Dorey} was reproduced \cite{BV}(a different calculation that is valid at weak coupling had similar results for other sets of bound states \cite{HO}).
 Moreover, these techniques generalize very easily to orbifold setups
\cite{BCorr,BCott}, where one realizes that the set of commuting matrices up to gauge equivalence gets replaced by the moduli space of vacua of the corresponding field theory.
Finite $N$ effects can also be explored numerically \cite{BCott2} and a numerical simulation of three point functions is underway \cite{BCL}. Recent studies of SYM at finite temperature have also revealed a similar structure \cite{Allt}. 

The idea of this paper is to show that one can extend this procedure for other more complicated field theories: all the superconformal field theories that are dual to $AdS_5\times X$ solutions
of type IIB string theory,  for $X$  a Sasaki-Einstein manifold (the so called non-spherical horizons \cite{MP}). Here we consider the case of only a Freund-Rubin ansatz\cite{FR}.  

In this paper I will propose a way to address these field theories in a unified framework by doing calculations in field theory at strong coupling from first principles.  
Obviously we need to find some suitable approximation of the strong coupling dynamics that lets one proceed to do calculations. 

The main result of this paper is that the dual states to chiral ring under the operator-state correspondence can be described by a set of classical configurations of the field theory compactified on $S^3\times \BR$. These configurations are in one to one correspondence withe the classical moduli space of vacua of the superconformal field theory in flat space.

The idea for solving these theories at strong coupling is that the field configurations that explore the moduli space of vacua dominate the low energy dynamics. Indeed, this is the minimal set that allows a complete description of the chiral ring. We will argue
that we have a conformal bootstrap that extends the classical dynamics associated to the chiral ring
states to a complete characterization of the low energy 
effective dynamics of the field theory at strong coupling.

 The moduli spaces of these theories are given essentially by $N$-D branes on
the Calabi-Yau cone over $X$, which we will call $V$. Essentially, we find a toy model problem of $N$ particles moving on $V$. We will find that if one quantizes the system with techniques similar to those found in ${\cal N}=4$ SYM \cite{Droplet} one also gets a description in terms of a gas of particles with effective repulsive interactions. The saddle point of this gas will locate the particles far away from the tip of the cone
of $V$, so one is in a region of field space where the transverse variables to the  moduli space become heavy and some of them can be treated perturbatively in effective field theory.

The argument will be a self-consistent argument. I will describe carefully in what sense the moduli space dynamics dominates the low energy effective field theory. If one makes this assumption, one can solve the reduced dynamics.  The idea is to show that after one solves the reduced dynamics, the transverse directions to moduli space do satisfy all the
assumed conditions, and then I will proceed to do some of the perturbative calculations.

The moduli spaces associated to these field theories are simple. They roughly correspond to $N$ D-particles on $V$. Thus one finds that the system also reduces to a gas of particles in $V$. There are two types of singularities in the dynamics: when one particle goes to the tip of the cone in $V$, and  when two particles in $V$ coincide. The first degeneration is how one defines the quiver gauge theory for a single brane, and the naive vacuum of the theory places all of the D-brane at the origin rendering the theory incalculable.

For the second type of degenerations, the dynamics has an enhanced gauge symmetry, and the low energy dynamics near one of these degenerations is identical to ${\cal N}=4 $ SYM. This is expected from the axioms of D-geometry \cite{Douglas,DKO}. In the case of ${\cal N}=4 $ SYM there is an effective repulsion of the eigenvalues induced by a change of variable Jacobian measure, and this should be true for all the CFT's that we are studying. This repulsion moves the D-particles away from the origin, and places them at finite distance on the moduli space. The particles are going to be much closer to each other on average than to the tip of $V$.
The transverse directions to moduli space can be thought of as open string-bits stretching between D-particles. The masses of these string bits are (at least locally) proportional to distances between the D-particles (as required by the axioms of D-geometry), so the lightest string bits correspond to open strings 
between nearby D-particles. When two D-particles coincide we get an enhanced local 
${\cal N}=4$ SYM dynamics, with some effective ${\cal N}=4 $ SYM coupling constant $g_{eff}$. The effective t' Hooft coupling \cite{'tH} will be $g^2_{eff}N= \lambda$, and $\lambda$ controls the mass of the string bits relative to the radius of $S^3$ space on which the field theory is compactified. For large $\lambda$ it will be found that the string bits are heavy and can be integrated out systematically. 

The dynamics that will dominate is the dynamics of the moduli space of vacua plus the
most likely degenerations. This is locally the same as ${\cal N}=4$ SYM, so we find a universal description of the dynamics of all these field theories that is very similar to the theory where these problems were already solved.

This universality of dynamics is responsible for guaranteeing that all the dual geometries will be
associated to a type IIB supergravity background. Also the same mechanisms that lead to a metric on the $S^5$ that can be measured in string units will work in this general case as well. 

Given the formalism one can then test limits where the string spectrum is known directly in field theory.
In particular, we will find that we are able to match the BMN \cite{BMN} limit spectrum of strings for a non-trivial field theory: the Klebanov-Witten theory. This has been studied in
the gravity theory previously \cite{IKM}.

The paper is organized as follows.

In section section \ref{sec:SCA} we review the ${\cal N}=1$ superconformal algebra and 
the derivation of the BPS inequality between the energy and R-charge of representations and it's relation to unitarity of the quantum field theory. We argue also why in a classical treatment of the theory this inequality is automatically satisfied.

In section \ref{sec:class} we derive various aspects of classical supersymmetric conformal 
field theories. In particular, we derive the scaling properties of the Kahler potential that are required for superconformal invariance, as well as the conformal coupling of the scalars to a
background metric. We find that this conformal coupling is proportional to the Kahler potential.

Next, in section \ref{sec:BPS}, we describe the classical solutions of the supersymmetric
field theory compactified on $S^3\times \BR$ that solve the BPS bound. We find that these solutions are in one to one correspondence with points in the moduli space of vacua of the theory in flat space.

In section \ref{sec:Dgeom} we discuss these moduli spaces in detail for the
theories that are dual to $AdS_5\times X$ geometries. In particular, we show how 
given the action of theory one can find that these moduli spaces and the effective dynamics near them follows the axioms of D-geometry \cite{Douglas}: the moduli space can be described by $N$
D-particles on a background geometry, and the excitations can be understood as open strings stretching between the D-particles, with masses of the open strings that are to first order proportional to the distances between the corresponding D-particles.

The essential argument of this paper is then presented in section \ref{sec:SC}. This shows
the consistency of the strong coupling expansion around moduli space configurations and in particular it argues that the wave function of the effective dynamics in the moduli space of vacua localizes near configurations where all the off-diagonal modes (open strings in the D-brane picture) are very heavy.

In section \ref{sec:emg} the notion of emergent geometry is analyzed in these quantum field theories. In particular, a description of how locality in $X$ arises from energetic considerations in the dynamics of the ground state is given. Also, a calculation of the spectrum of states that are dual to strings in the BMN limit of the Klebanov-Witten theory is presented. This is a non-BPS test calculation in a quantum field theory that has no gaussian fixed point.

Finally, a discussion of the issues that still need to be resolved will be presented at the end in section \ref{sec:disc}.

\section{The $N=1$ superconformal algebra} \label{sec:SCA}

In this paper we will be dealing with ${\cal N}=1 $ SCFT in four dimensions. 
It is useful to  collect some basic properties of the superconformal algebra for further use later on.

The algebra generators can be classified according to their weight under rescalings. We will call $\Delta$ the generator of dilatations around a fixed origin. The generators are given by the following diamond
\begin{equation}
\begin{matrix} && K_\mu &&\\
& S_{\alpha} && \bar S_{\dot \alpha}&\\
\Delta && M_{\mu\nu} && R\\
& Q_\alpha && \bar Q_{\dot \alpha} &\\
&& P_\mu &&
\end{matrix}
\end{equation}
 The generators $Q, \bar Q$ are the usual supersymmetry generators and $P_\mu$ is the generator of translations in flat space. These have dimension $1/2, 1/2, 1$ respectively. 
 The generator of dilatations is $\Delta$, the generators of rotations are $M_{\mu\nu}$, of dimension zero. The conformal group also has special conformal transformations, whose generators are $K_\mu$, these are of dimension $-1$. The commutator of $K$ and $Q$ requires the presence of some other spinor generators of dimension $-1/2$. These are the $S_\alpha$ and $\bar S_{\dot \alpha}$, the special conformal transformations. Closure of the algebra then requires a new generator $R$, the so called $R$- charge, that commutes with the Lorentz group (and the conformal group). The algebra is written in detail in \cite{West}.
 
 The algebra is graded with respect to the dimension of the generators. This lets us determine some structure constants by standard dimensional analysis and Lorentz invariance. For example, up to normalization constants we must have that 
 \begin{equation}
 \{ S_\alpha, \bar S_{\dot\beta}\} \sim \sigma^{\mu}_{\alpha \dot\beta} K_\mu
 \end{equation}
 Also
 \begin{equation}
 \{S_\alpha, \bar Q_{\dot\alpha} \}= 0
 \end{equation}
 because there is no vector representation of the Lorentz group as generators of dimension equal to zero.
 
 The most important commutator for our purposes is
 \begin{eqnarray}
 \{ Q_\alpha, S_\beta\}\sim (\Delta + J) \epsilon_{\alpha\beta}+ M_{\mu\nu} \sigma^{\mu\nu}\!_{\alpha\beta}\\
  \{ \bar Q_{\dot \alpha},\bar S_{\dot\beta}\}\sim (\Delta - J) \epsilon_{\dot\alpha\dot\beta}+ M_{\mu\nu} \bar\sigma^{\mu\nu}\!_{\dot\alpha\dot\beta}
 \end{eqnarray}
 which is how the $R$-charge appears (here it is represented by a generator $J$). There is an issue of the choice of normalization of the R-charge. We will use the convention where 
 the coefficients above are always $\Delta\pm J$. This differs from the usual normalization where $J=3/2R$. With the normalization above, a free chiral scalar superfield has dimension and $J$ charge equal to one, and the superspace $\theta$ variables have $J$- charge equal to $3/2$.
 
 The algebra as described above has various real forms that are important. For flat Minkowsky space we have that $P$ is self-adjoint,  and $Q$ is the complex conjugate of $\bar Q$. A similar result holds for $S, \bar S$  and $K$. A typical scattering experiment (as given in perturbation theory), would start with particles at some fixed momentum $p_\mu$ and with some spin labels, and we would have a similar description of the out-state. Since all particles (to the extent that one can talk about single particle states)  are massless in conformal field theories, there are severe infrared divergences in standard S-matrix calculations, mostly from soft collinear emission.
 
 A second possibility to understand the representations of the algebra occurs when we go to an Euclidean field theory setup. In that case, local insertions of operators are classified by representations of the conformal group. An operator $\CO$ is called primary if it is annihilated by all the $K$, namely, if
 \begin{equation}
 [K_\mu, \CO]= 0
 \end{equation}
An operator is called superprimary if it is annihilated by the $S,\bar S$ operators. This is, if
\begin{equation}
[S, \CO]_{\pm} = [\bar S, \CO]_{\pm}=0 
\end{equation}
the graded commutators with $S, \bar S$ vanish.
A superprimary operator is automatically a primary operator (this can be checked by using the Jacobi identity). The set of primary operators can be classified according to their dimension,  $R$-charge and spin, so that they satisfy
\begin{equation}
[\Delta,\CO] \sim \Delta_{\CO}\CO
\end{equation}
here $\Delta_{\CO}$ is the dimension of $\CO$, and similarly for $J$, $M$. The reason for this is that acting with $\Delta, J, M$ preserves the property of being a super primary operator.

In conformal field theories it is well known that there exists an operator-state correspondence. This states that if we take the punctured $\BR^4$, where we substract the origin, and we consider writing the metric of flat $\BR^4$ in spherical coordinates, we have that
\begin{equation}
ds^2 \sim r^2 \left( \frac{dr^2}{r^2}+ d\Omega^2_3\right) \simeq d\tau^2+ d\Omega_3^2
\end{equation}
the metric is conformally equivalent to that of the three sphere times the real line, where we have used $\tau= \log(r)$. Thus the  euclidean field theory compactified on $\BR^4/0$ and $S^3\times \BR$ are related. The origin in $\BR^4$ gets sent to $\tau\to -\infty$, so we change the insertion of an operator at $r=0$ (this is the reason why we puncture the plane) for a boundary condition in the infinite past (boundary conditions in Euclidean field theories
correspond to states in regular quantum field theories).

After an analytic continuation where we set $ \tau= i t$, we have a standard quantum field theory with a real time. The states in the euclidean field theory and the real  time field theory are the same. The generator of dilatations $\Delta$ corresponds to the vector field $r\partial_r\sim \partial_\tau\sim -i \partial_t$, so it is the generator of time evolution in the quantum field theory on $S^3\times \BR$, this is, we need to identify the operator $\Delta$ with the Hamiltonian of the field theory on $S^3\times \BR$. 

Unitarity of the Hilbert space and requiring that $H$ is self-adjoint
requires that in the field theory on $S^3\times \BR$
that $Q$ is the adjoint of $S$  (more precisely, $Q_{\alpha}$ is the adjoint of $S^{\alpha}$), and $\bar Q$ is the adjoint of $\bar S$, while $K$ is the adjoint of $P$. The commutators with $H=\Delta$ make the $Q,S$ act as raising/lowering operators for the energy respectively.

This is a different real form of the superconformal algebras than the one associated to flat Minkowsky space. It is with respect to this real form that we usually speak about unitary representations of the conformal group.

Since standard field theories usually have a positive spectrum of energies, the operators $K$ and $S,\bar S$ act by lowering the energy of a given state by a fixed amount. Eventually, we always end up with states that 
are annihilated by $S, \bar S$. These are lowest weight states with respect to the conformal algebra. These states are the dual states to superconformal primaries $\CO$, so we call them the same way $\ket{\CO}$.

We can raise the energy of a state by acting on it with $Q,\bar Q$. We obtain this way a list of 
new states that are required to be there by the conformal group once we have the primary states. These states are called descendants. Unitarity requires that all linear combinations of descendants have a non-negative
norm. This leads to BPS-type inequalities. 

For example, take the collection of states $Q_{\alpha}\ket{\CO}$, where $\CO$ is superprimary and therefore annihilated by the $S$.  For its norm to be non-negative, it requires 
that
\begin{equation}
|\bar Q_{\dot \alpha}\ket{\CO}|^2= \bra{\CO} S^{\dot\alpha} Q_{\dot\alpha} \ket{\CO}
=\bra{\CO} \{S^{\dot\alpha}, Q_{\dot\alpha}\} \ket{\CO}\sim \bra {\CO}
\Delta -J+ M^{\dot\alpha}\!_{\dot\alpha}\ket{\CO}\geq 0 
\end{equation}
This tells us that the dimension has to be larger than the R-charge plus part of the spin (remember that the rotation group $SO(4)\sim SU(2)\times SU(2)$ requires two spin quantum numbers, left and right).
 If we sum over the spin index above, we obtain that 
 \begin{equation}
 \Delta\geq J \label{eq:BPS}
 \end{equation} 
 because the $M_{\alpha\beta}$ are symmetric and the raising index operator is $\epsilon^{\alpha\beta}$ that is antisymmetric. Primary superconformal states that saturate the equality are called BPS states. Our assumption above is that we have already chosen conformal primary states.

A similar argument follows from considering descendants $Q\ket\CO$, so that we find  
\begin{equation}
\Delta \geq |J| 
\end{equation}
Primary states such that $\Delta_{\CO} =- J_{\CO}$ will be called anti-BPS.

It is clear that if we take descendants (only under the conformal group), then 
we change the value of $\Delta$, but not that of $J$ (the generator $J$ commutes with the Lorentz group generators). Thus we will not be able to saturate the BPS equality with descendants. We find this way that the states that saturate the BPS bound are always primaries.
More work is required to show that they are always superprimaries.

If we take into account
only primary fields and descendants under the conformal group, the representations of the conformal group are 
in general very simple: we just have local primary operators and  derivatives acting on primaries and their rotations. The representations can be reconstructed from the primary fields (there are no null states appearing in the list of descendants unless one has a free field).

Furthermore, a superconformal representation decomposes into finitely many primaries with respect to the conformal group. This is because up to taking descendants, we have that $Q^2=0, \bar Q^2=0$, $\{Q, \bar Q\}= P \sim 0$, so the $Q$ anticommute with each other and are 
nilpotent. The list of distinct monomials in $Q$, $\bar Q$ is finite and this translates into finitely many conformal primaries. This is similar to the fact that supersymmetry representations in flat space give rise to finitely many particles in a supermultiplet. 

Another useful point of view on BPS states is as follows: we consider a general superfield
multiplet of operators
\begin{equation}
\Phi = \phi + \theta\psi +\bar\theta \bar \psi +\theta\bar \theta V +\theta^2 F +\bar\theta^2 \bar F+\dots
\end{equation}
The superfield is a chiral superfield  if $\bar D\phi=0$. In superspace we have schematically (see \cite{WB} for details on superspace notation)
\begin{eqnarray}
\bar D\sim \partial_{\bar \theta} + \theta \sigma^\mu\partial_\mu\\
\bar Q\sim \partial_{\bar\theta} - \theta\sigma^\mu\partial_\mu
\end{eqnarray}
and if we take the expansion of the superspace differential equation (the chiral superfield constraint) $\bar D \Phi=0$ we find that
\begin{equation}
\bar \psi +\theta V + \bar \theta F +\dots + \theta \partial \phi +\dots=0
\end{equation}
so that $\bar \psi=0$ and that $V= -\partial \phi\sim [P,\phi]$, etc. This is, various of the components of the superfields are either zero (null, or of zero norm if we talk about the corresponding state), or they are descendants of other components. 

 Similarly, we can act with $\bar Q$ on $\Phi$ to find that
\begin{equation}
\bar Q \Phi \sim \bar \psi+ \bar\theta (V-\partial\phi)+\dots 
\end{equation}
and if we use the chiral constraints described above, we get that $[\bar Q,\phi]=0$.
This is, the lowest component of a chiral field is annihilated by some of the 
supersymmetries. In the state language, after using the operator state correspondence,  we would have $\bar Q \ket{\phi}=0$. 

Together with being primary, this condition implies that we are saturating the BPS bound. Thus chiral field superprimaries are BPS operators. The asssociated representations of the superconformal group are small in the sense that some primaries in the decomposition of the supermultiplet into conformal multiplets are absent. Because BPS representations are smaller (some descendants are null), they must combine with other representations to become non-BPS and this property allows for the existence of an index of BPS states to be defined \cite{index}.

A particularly important set of operators is the chiral ring. This is the cohomology of $\bar D$: the collection of 
superfields such that $\bar D \Phi=0$, but $\Phi$ can not be written as a $\bar D $
derivative acting on something else. In superconformal field theories, the set of (half) BPS operators is identical to the set of lowest component operators of chiral ring elements.  
The chiral ring also has powerful non-renormalization theorems attached to them, so it is protected by supersymmetry in going from weak to strong coupling, except perhaps for 
non-perturbative corrections to some chiral ring relations, that involve gaugino condensation in the vacuum  \cite{S} ( see also \cite{CDSW} for an in-depth description of properties of the chiral ring). 
These corrections are naively expected to vanish in superconformal field theories, as there is no obvious
QCD confining scale that could generate non-vanishing gaugino condensates. One can show that sometimes some of the branches of the  moduli space of vacua are lifted
by quantum corrections \cite{Brun} in theories that are geometrically engineered. These effects seem to only happen in D-brane constructions where one can separate `fractional branes' away from the origin, this is, for a case of branes near a Calabi-Yau singularity that is not isolated. We will not consider this case in this paper.

The importance of the chiral ring is that it is independent of the Kahler potential and can be calculated non-perturbatively. Operator product expansions of chiral ring operators are non-singular, and this is what gives them a ring structure: we keep the non-vanishing terms in the OPE when operators collide. If we have a superconformal field theory, each chiral ring operator has an R-charge and the dimension of the operator is equal to the R-charge. Products of chiral ring objects that are non-vanishing in the chiral ring have an R-charge that is equal to the sum of the charges of the elements in the product. Thus, one can compute the dimension of composite product operators classically: it is additive on the constituents. Moreover, the
chiral ring vevs serve as order parameters that classify the different supersymmetric vacua of a theory. In a conformal field theory, their vacuum expectation values parametrize the moduli space of vacua of the field theory. 

In our study we will be interested in studying the dual states to chiral ring operators in a conformal field theory. In particular we want to understand if one can establish a more precise relation between the moduli space of vacua of the theory and the chiral ring 
states. 
The BPS inequality will be our guide in this respect.  Since the inequality follows from unitarity, 
if we have a classical theory that can be quantized, then the usual quantization rules produce a unitary quantum field theory automatically. Thus, the BPS inequality on the phase space of the field theory must be realized as a classical inequality between two classical observables on the set of all configurations: the energy and the R-charge. We will try to understand what are the implications for classical physics of satisfying this inequality. This will be our starting point
to explore the field theories at strong coupling. The basic constructions will setup the problem of strong coupling so that the chiral ring states are for the most part automatically accounted for by the effective dynamics.

\section{Classical Preliminaries}\label{sec:class}

We will assume that we have a classical supersymmetric gauge field theory in four dimensions
that is superconformally invariant. The extent to which one can apply classical reasoning 
will be discussed further on. For the meantime, we will assume that that the theory has been written in terms of a collection of fundamental chiral scalar fields $\phi^i$, whose
classical dimension is $\gamma_i$, and whose $R$ charges are proportional to 
$\gamma_i$. This usually follows from the representation theory of the superconformal algebra, as described before. 
We will also have gauge fields, $A_\mu$, represented by  vector multiplet superfields. 
In order for covariant derivatives $\nabla_\mu \phi$ to have canonical dimension uniformly, $A_\mu$ has to have canonical dimension equal to one. 

We will assume because of this that the action for the gauge fields is the standard SYM action, with a coupling constant $g_{YM}$ that does not depend on the $\phi$. This choice seems to be unique: the gauge coupling constant must be holomorphic. Also, in the classical theory, standard dimensional analysis applies, thus the coupling constant must be homogeneous of degree zero with respect to the $R$-charge.
Any such non-trivial quantity will be in the form of a fraction of holomorphic fields, and will be singular at various finite values of the fields. To the extent that we are working with classical fields, 
such singularities are unphysical and should be absent. They would indicate that we 
have an incomplete dynamics. Another point of view that one can have here is that if the theory is superconformal then the beta functions of the coupling constants are zero, and therefore the coupling constants should not depend on the fields (the fields would parametrize spontaneous breaking of conformal invariance). The reasoning here is classical and applies to situaitons hwere the gauge coupling constants can be taken top be small, thus some conformal field theories thatcan not be described classically (like the ones associated to an Argyres-Douglas fixed point \cite{AD}) are not covered in this setup.

For the scalars, we will have a Kahler potential $K(\phi, \bar \phi)$
and other than requiring compatibility with superconformal invariance, we will not place any additional restrictions on $K$. We will also have a polynomial superpotential $W(\phi)$, which is homogeneous with respect to the R-charge of the system. 
Indeed, all superconformal field theories in four dimensions that have been constructed have this property (polynomial superpotential), so it is a reasonable assumption to make. This will  also prevent singularities of $W$ near the origin of field space $\phi=0$.

This is the starting point of our analysis. We will show various properties of $K$ that follow from superconformal invariance. We will also show that the conformal coupling of the scalars to the
background metric requires a term in the action of the form
\begin{equation}
a \int \sqrt{-g} R K(\phi, \bar\phi)
\end{equation}
where $R$ is the Ricci scalar of the background metric, and the value of $a$ can be calculated explicitly.

First, we will show that $K$ is homogeneous with respect to the holomorphic fields $\phi$ and $\bar\phi$ respectively. 

In order for the field theory to be conformally invariant, the theory must be compatible with
an $R$-charge symmetry and $K$ should have a scaling property so that the Lagrangian density is classically scale invariant.
 This should be a symmetry of the Kahler potential that is realized linearly, because when the fields vanish we are at a point where superconformal invariance is not spontaneously broken.
 
 It is easy to show that if the fields $\phi$ have R-charge $\gamma_i$, and if we require $K$
 to be invariant with respect to R-charge rotations of the $\phi$, then the kinetic term of the lagrangian is  
 going to be R-charge invariant, as it is given by
 \begin{equation}
 \int d^4 \theta K
 \end{equation}
This condition, that $K$ is invariant, translates into the following differential equation
\begin{equation}
\sum _i\gamma_i \phi^i K,_i - \gamma_i \bar\phi^i K,_{\bar i}=0\label{eq:Rinv}
\end{equation}
where we are summing over all fields.

Similarly, under an infinitesimal dilatation rescaling of the metric, we have 
\begin{equation}
g_{\mu\nu} \sim \exp( -2\sigma) g_{\mu\nu}\label{eq:mscaling}
\end{equation}
for constant $\sigma$, then 
\begin{equation}
\phi^i\to \exp(\gamma_i \sigma) \phi \label{eq:sscaling}
\end{equation}
 and 
it's complex conjugate as well
\begin{equation}
\bar \phi^i\to \exp(\gamma_i \sigma)\phi,
\end{equation} 
where in the above $\sigma$ is a constant parameter. In order for $\int d^4 x \int d^4\theta K$ to be invariant we must have that $K$ is of classical dimension equal to $2$ (this is the same result as in free field theory). Thus, we must have the following differential equation
\begin{equation}
\sum_i \gamma_i \phi^i K,_{i}+\gamma_i{\bar \phi^i} K,_{\bar i} = 2K \label{eq:Sinv}
\end{equation}
Combining the equations  \ref{eq:Rinv} and \ref{eq:Sinv} we find that
\begin{eqnarray}
\sum_i \gamma_i \phi^i K,_i &=& K \label{eq:hom}\\
\sum_i \gamma_i \bar\phi^i K,_{\bar i} &=& K
\end{eqnarray}
This is, $K$ is a homogeneous function both with respect to the chiral and the anti-chiral fields separately, where each chiral field is weighed by its R-charge. This is a generalization of the scaling properties for the standard Kahler potential for a collection of free fields $K \sim \sum \phi\bar\phi$, where the field $\phi$ has dimension equal to one. Extra relations can be obtained by taking derivatives. For example
\begin{equation}
\sum_i \gamma_i \phi^i K,_{i\bar j} = K,_{\bar j} \label{eq:homder}
\end{equation}
follows because $\phi$ does not depend as a variable on the $\bar\phi$ coordinates. We will make use of such relations many times over.

Conformal invariance should be understood in terms of local scale invariance with respect to a background field metric $g$ that also transforms: this is, the conformal field theory should only couple to the conformal class of the metric, and not to the full metric degrees of freedom (this point was heavily emphasized in \cite{Wads}). We want to understand this at the classical level.
If we consider the kinetic term of the theory, and we do local transformations as in \ref{eq:mscaling} and \ref{eq:sscaling}, we find that the kinetic term is not conformally invariant on its own. Indeed, the kinetic term is given by
\begin{equation}
S(g,\phi) = \int d^4 x \sqrt{-g}  K,_{i\bar i } g^{\mu\nu} \partial_\mu \phi^i  \partial_\nu \bar \phi^{ i}
\end{equation}
and this transforms to first order as
\begin{equation}
S(g e^{-2\sigma}, e^{\gamma_i\sigma}\phi) \sim
S(g,\phi) + \int d^4 x \sum_{i}  K,_{i\bar i} g^{\mu\nu} (\partial_\nu \phi^i \gamma_{\bar i} \bar\phi^i +\partial_\nu\bar \phi^i \gamma_i \phi^i ) \partial_\mu\sigma
\end{equation}
It is easy to use the homogeneity relations \ref{eq:hom} to show that the variation is proportional to
\begin{equation}
\delta S \sim \int d^4 x \partial_\mu K g^{\mu\nu} \sqrt{-g} \partial_\nu \sigma
\end{equation}
We can now integrate by parts, and using  covariant derivatives, we find that
\begin{equation}
\delta S \sim - \int d^4 x K \sqrt{-g} \nabla^2 \sigma
\end{equation}
where we are using the Laplacian in the background metric $g$.

We find this way that the kinetic term alone is not strictly conformally invariant. We can add a 
non-minimal coupling to the metric, of the form
\begin{equation}
a \int d^4 x \sqrt{-g} R K
\end{equation}
where $R$ is the Ricci scalar of the curvature $g$. It is well known that under infinitesimal conformal rescalings of the metric $R\to R+ b \nabla^2 \sigma$, where $b$ depends on the dimensionality of spacetime and various sign conventions (see \cite{Wald} appendix D for example). The non-minimal coupling has the correct dimension classically.
The coefficient $a$ should be balanced so that the full lagrangian is classically conformally invariant, and it is independent of the details of the field theory. The coefficient $a$ is the same number that appears in the conformal coupling that is required for coupling free fields conformally 
to a background metric.

These issues only affect the bosonic scalar degrees of freedom. For the gauge fields, the classical Yang Mills action is already conformally invariant, and $A$ does not transform under a metric rescaling. This is also expressed sometimes by saying that the critical dimension for the Yang Mills action is $d=4$.

Our conventions are such that if we take a units 3-sphere and we write the
field theory on $S^3\times \BR$, then we have that the kinetic term plus the conformal 
coupling to the metric add up to
\begin{equation}
S_{kin}\sim \int_{ S^3} \int d t \left\{ K,_{i\bar i}\left[ ( D_t\phi^i D_t \bar \phi^i) - \nabla\phi^i\cdot \nabla\bar \phi^i\right]-  K\right\} \label{eq:confcoup}
\end{equation}
This can be checked for the special case of ${\cal N}=4$ SYM, and we find this identical form.

With these conditions on the metric, the kinetic Hamiltonian for the scalar components of the field theory is given by
\begin{equation}
 H_{kin} \sim \int_{S^3} \left\{ K,_{i\bar i}\left[ ( D_t\phi^i D_t \bar \phi^i) + \nabla\phi^i\cdot \nabla\bar \phi^i\right]+ K\right\}
\end{equation}
where we use the $\nabla \phi$ to indicate gradients along the sphere.
The total Hamiltonian is
\begin{equation}
 H \sim \int_{S^3} \left\{ K,_{i\bar i}\left[ ( D_t\phi^i D_t \bar \phi^i) + \nabla\phi^i\cdot \nabla\bar \phi^i\right]+ K\right\}+ H_{YM} + H_{f}+ \hbox{F, D-terms}\label{eq:clham}
\end{equation}
where $H_f$ is the contribution to the energy from the fermions, and $H_{YM}$ is the standard Hamiltonian for a Yang Mills field on the sphere. The $F$ and $D$-terms result from eliminating the auxiliary $F$ and $D$ terms in the superspace formulation of the theory. They are positive definite (a sum of squares).

Also, the R-charge generator, that can be computed using Noether's procedure, is given by
\begin{eqnarray}
Q_R&\sim& \int \frac{\delta S}{\delta{\dot\phi}} \delta_Q\phi\\
&\sim&\int _{S^3} \sum\left(K,_{i\bar i}  (-i\gamma_i \phi^i) D_t \bar\phi^i
-K,_{i\bar i} D_t\phi^i (-i \gamma_{\bar i} \bar\phi^i) \right)\\
&=& -i\int_{S^3} \left( \sum_i K,_{\bar i} D_t \bar\phi^i-K,_i D_t\phi^i \right)\label{eq:Q}
\end{eqnarray}
where to obtain the last line we have made use of the homogeneity equations that 
$K$ satisfies, in particular we have used equation \ref{eq:homder} and its complex conjugate.
We are using conventions where $\delta_Q\phi= -i\gamma_\phi\phi$ (this is well defined up to the sign of $Q$).

For gauge theories on $S^3\times \BR$ it is often convenient to take the gauge condition $A_0=0$. For that case, covariant time derivatives become ordinary time derivatives. We supplement 
the equations of motion in this case by adding the gauge constraints (the equation of motion of $A_0$).  It is very useful to work in the Hamiltonian formulation, a setup that respects all the isometries of the system. In the Hamiltonian language, we can think of a field theory configuration space as an infinite-dimensional phase space with the gauge constraints commuting with the Hamiltonian vector field.

\section{Classical BPS solutions}\label{sec:BPS}

We will now show that classical BPS solutions of the field theory compactified on $S^3$ - those that correspond to states dual to the chiral ring- are in exact correspondence with the moduli space of vacua of the theory. The idea in this case is to explore the structure of BPS states in
the same spirit as was done in \cite{Droplet}.

Indeed, consider an element of the chiral ring $\CO$, of some dimension $\gamma_{\CO}$ (this is well defined at a superconformal ground state).
 These are chiral operators $\CO$ (made only out of polynomials in the chiral elementary fields). In particular, $\CO$ can not be a descendant of a chiral primary.

The lowest component of a chiral ring elements are special. Their conformal dimension is equal to their $R$ charge (in our normalization this is called $J$). This is the same argument used in equation \ref{eq:BPS}. Since the $R$ charge is additive, we find that the dimension of $\CO$ is the sum of the dimension of its constituent fields: this is, the classical counting holds.

Such elements of the chiral  
 ring give rise to short representations of supersymmetry, and short representations are usually called  BPS. We want to now use the operator state correspondence to describe the possible classical field configurations on $S^3$ that are dual to such operators. Any such operator $\CO$, will give rise via the operator state correspondence to a dual state $\ket{\CO}$ such that
 \begin{equation}
 H \ket{\CO} = Q \ket{\CO}= \Delta_{\CO} \ket{\CO} \label{eq:BPSs}
 \end{equation}
 where $\Delta_{\CO}$ is the dimension of $\CO$, and $Q$ is the conserved charge operator in the Hamioltonian system. Linear combinations of these states also satisfy $H=Q$. Thus a classical state $\ket s$, which can be understood in  general as a coherent state in a quantum mechanical system, will be described by such a linear combination of states whose energies 
are very near each other.

Such a classical state is characterized by a classical trajectory of the dynamical fields $\phi(t)$ (we will include the gauge fields here, but not the fermions, as the fermions can not have classical vevs). 
These trajectories are calculated by using 
\begin{equation}
\phi(t) \sim \bra {s(t)} \hat \phi \ket { s(t)}
\end{equation}
where the state $s$ is evolved according to the usual Schr\"odinger equation, and $\hat \phi$ is the corresponding operator in the quantum mechanics (we are working in the Schrodinger picture, so $\hat\phi$ is time independent). One can show using equation \ref{eq:BPSs} that for these states it must be true that 
\begin{eqnarray}
\bra {s(t)}[ \hat \phi,H] \ket { s(t)} &=& \bra {s(t)}[ \hat \phi,Q] \ket { s(t)}= -  R_\phi
\phi(t)\\
 &\sim& i \dot \phi(t)
\end{eqnarray}
The right hand side uses the fact that the fields can be chosen to have definite $R$-charge. 
Otherwise we write it in terms of the vector field associated to $Q$ in the Hamiltonian system.
The left hand side in the classical limit becomes the Poisson bracket of $H$ and $\phi$ evaluated on the state, and it describes the time evolution of the system. 
If this is true, we find that the corresponding classical states should have very simple trajectories. 

We will now argue using classical physics that this quantum mechanical intuition is correct. The idea is that the BPS function $H-Q$ is positive on the configuration space of the dynamical system, and it acquires a global minimum for BPS states. This inequality is a consequence of unitarity in the quantum theory.
If standard quantization always produces a unitary theory, the inequality must be a true statement in the classical phase space before quantization.

Since $H,Q$ have vanishing Poisson brackets (Q is a conserved charge after all), the set 
$H-Q=0$ is preserved by the evolution according to the two Hamiltonian vector fields associated to $H$ and $Q$. Since $H-Q$ acquires a global minimum, it is locally quadratic (or higher order) in local variables of the dynamical system when we Taylor expand around such a minimum. Because of this, the Hamiltonian vector field associated to $H-Q$ must vanish on this locus : this is the familiar fact that minima - or any extrema for that matter- of a Hamiltonian function on phase space give rise to 
static (time independent) solutions of the equations of motion. Thus, for BPS states we have that the Hamiltonian evolution according to $H$ is the same as the Hamiltonian evolution according to $Q$, because for these states the evolution according to the vector field $H-Q$ vanishes. This is the classical argument that shows that the quantum intuition is correct.

Let us now calculate $Q$ and $H$ for these types of classical trajectories. This will let us completely characterize the set of solutions which saturate the BPS inequality.

We easily find that (in the gauge $A_0=0$) that the BPS classical states correspond to an $J$ charge given by
\begin{equation}
Q=-i \int_{S^3} \left( \sum_i K,_{\bar i} \partial_t \bar\phi^i-K,_i \partial_t\phi^i \right)=
2 \int_{S^3} K
\end{equation}
where the homogeneity relation of $K$ is used again, and we are using $\dot\phi=i R_\phi \phi$, the simplified equations of motion and substituting in \ref{eq:Q}.

Moreover, we have that for the energy on the $S^3$
\begin{equation}
H = \int_{S^3} 2K + K_{i\bar i} \nabla \phi^i \nabla \bar \phi^{\bar i}+ H_{YM} + \hbox{(F+D-terms)}
\end{equation}
where again the kinetic term is computed using the homogeneity relations.

We find that $H-Q$ vanishes only if $\nabla\phi= H_{YM}= F = D = 0$, as we find a sum of squares involving all these terms. Let us analyze these in turn. Having the energy in the YM field vanish, implies that the field strength vanishes. Thus the gauge connection is flat. Since $S^3$ is
simply connected, there are no non-trivial Wilson lines and we find that $A$ vanishes 
identically (this usually depends on a gauge choice, but it is compatible with the gauge $A_0=0$ that we have used so far).

Also, having $\nabla\phi=0$ implies that $\phi$ is (covariantly) constant on the sphere. 
Moreover, having the condition $F=D=0$ tells us that the corresponding constant configuration
is associated to a solution of the F and D-constraints that characterize supersymmetric vacua 
in flat space. Because of gauge invariance, only the equivalence class of the solution matters, so we find that the set of such configurations is in one to one correspondence with the points in the moduli space of vacua of the superconformal field theory (this is after all, the set of solutions of the F, D constraints, modulo gauge transformations).
This proof is much more general than the one given in \cite{Droplet} that depended on having 
an oscillator description of the BPS states. The argument given above would work also in 
conformal field theories that are not marginal perturbations of a free field theory (let us say the
Klebanov-Witten CFT \cite{KW}).

We have thus found a direct relationship between classical BPS states of the field theory and the moduli space of vacua of the theory on flat space. It is known that these moduli spaces are complex manifolds with 
a Kahler metric. Because we have a simple relation between the velocities of the
fields and the fields themselves, we find that the moduli space is endowed with a natural 
non-degenerate Poisson structure: the natural induced Poisson structure on the manifold of initial conditions. It is easy to show that this ends up being proportional to the Kahler form
on the moduli space of vacua.

 The chiral ring is also related to such a moduli space: the chiral ring vevs coordinatize the moduli space of vacua in terms of complex variables, and they act as order parameters to distinguish the different vacua of the theory. With some mild assumptions (no nilpotent elements in the chiral ring), the holomorphic coordinate ring of the moduli space of vacua is identical to the chiral ring itself \footnote{In supersymmetric gauge theories one also has gaugino superfields $W_\alpha$ that are chiral. These are nilpotent and could in principle present problems. We do not have to worry about these: in the classical regime these are zero, because they always involve the fermion partners of the gauge fields as a lowest component, and they can be easily separated from the purely bosonic chiral superfields without trouble. The case that is more problematic is one where $\CO^n=0$, as in the case
of a superpotential for a scalar field of the form $W\sim \phi^{n+1}$. We will assume that we are working with the cases where the mild assumptions hold.}.

If we want to think of operators in the chiral ring as states, these operators furnish a set of holomorphic functions on the moduli space of vacua, and because of the canonical Poisson structure, they can each be interpreted as a holomorphic wave function on such a moduli space as a complex manifold. Considering that we found that the classical BPS states are also related to points in the moduli space of vacua, we find that the chiral ring is giving us a very precise holomorphic quantization of the moduli space of vacua. 

In general, this type of situation implies that the coherent states on the classical moduli space give an over-complete basis for the holomorphic quantization of such a space. Thus, one can argue that all elements of the chiral ring are related to quantum superpositions of classical BPS solutions: in essence, the classical analysis is enough to characterize the full quantum problem of the chiral ring, and makes it possible to apply a semiclassical analysis to various setups.
This can also be though of as having a magnetic field on the moduli space, where the magnetic field is proportional to the Kahler form, and further restricting the wave functions of a particle
moving in the moduli space to the lowest Landau level. This point of view generalizes 
ideas found in \cite{Btoy, Droplet} for this more general setup. Now, we will apply this insight to the dual field theories to type IIb strings on $AdS_5\times X$, for $X$ a Sasaki-Einstein manifold.

\section{The moduli space structure of $AdS_5\times X$ duals}\label{sec:Dgeom}

Now we have the basic tools we need to study some properties of gauge field theories that
are dual to $AdS_5\times X$, where $X$ is a Sasaki-Einstein manifold. The need for Sasaki-Einstein manifolds was explained in detail in \cite{MP}. 

We want to understand the dual field theories to such setups in as much detail as possible, and in essence, we want to derive the supergravity dual description from 
first principles: we want to solve the field theory and show that we obtain supergravity. 
This means that usual supergravity reasoning will only be used to guide the calculations, but we will not use any input from supergravity directly.

There are large collections of such pairs of field theories and dual geometries 
that are understood at a qualitative level: we know the matter content of the field theory and the structure of the superpotential (for example, branes at the tip of supersymmetric orbifolds $\BC^3/\Gamma$ \cite{DM}, or the field duals of branes at the tips of  $L^{p,q,r}$ cones \cite{HKW}). Also, the anomalous dimensions of the matter fields can be calculated using the a-maximization principle \cite{IW}. This permits to give a description of the chiral ring in detail.

From the gravitational point of view the chiral ring is special too. The R-charge is geometrically given by an isometry of $X$. The value of $J$ for a single particle moving in $AdS_5\times X$ is the angular momentum of the particle, as measured from this isometry, thus it is identified with momentum along a particular direction. The value of $\Delta$ is the energy of the particle in global $AdS$. The BPS bound $\Delta = J$ implies that the particle is massless in ten dimensions. Thus, in type IIB string theory it corresponds to a quantum of 
the gravity multiplet. These BPS objects can be calculated by solving for the spectrum of supergravity fluctuations of the geometry. Thus, the chiral ring objects, interpreted in the dual geometry, are given by pure supergravity solutions and they can be understood by solving ordinary partial differential equations. 
The techniques used to do this comparisons are purely based on holomorphy and
in this way one can only calculate properties of the chiral ring. It can be used as a check of the duality, but we would want to be able to make a much stronger statement by computing many non-BPS quantities as well and show that they have to match.

In order to do this, we need to exploit what we have learned so far about the chiral ring. We have found that in classical physics the chiral ring corresponds to trajectories that are related to points the moduli space of vacua, combined with the $R$-charge information . These moduli spaces are fairly well understood as complex manifolds, but we expect many additional properties related to the Kahler potential of these 
moduli spaces as well, that we need to determine. So far, we have only found out that
the Kahler potential of the CFT needs to have a scaling property. This still gives us a lot of freedom. We want to remove this freedom by requiring various properties. The most important one is that the theory should not just be classically superconformal, but that this property is true at the quantum level. We will assume that this will translate into the absence of
certain divergences in effective field theory (namely, these field theories should be finite in a similar sense to the${\cal N}=4 $ SYM theory and it's orbifolds).

The moduli spaces in AdS/CFT are supposed to follow the axioms of D-geometry \cite{Douglas} to some approximation.
The basic idea is that they represent $N$ point-like D-branes on some space, which for our purposes is a three-complex dimensional Calabi-Yau cone $V$. We will enumerate them
as follows

\begin{enumerate}
\item
The moduli space should be of the form 
\begin{equation}
{\cal M}\sim \hbox{Sym}^N V
\end{equation}
a symmetric product space: a collection of $N$ unordered points in $V$. The space has orbifold singularities where two (or more) of these unordered points coincide.

In this paper, we will assume that the cone has an isolated codimension three singularity and no codimension two singularities. Many Calabi-Yau cones have this property (for example the conifold and various orbifolds have this property).

\item 
The unbroken gauge  symmetry on the moduli space should be $U(1)^N$. When
we find ourselves in a typical singularity, we have exactly two D-branes on top of each other, and the
unbroken gauge symmetry is $U(2)\times U(1)^{n-2}$. If we stack more D-branes
on top of each other, the symmetry group is enhanced accordingly, producing various $U(M)$ gauge groups, following the usual rules for D-branes \cite{Pol}.

\item
To leading order it is also expected that the effective Kahler potential on the moduli space 
is a sum over the individual components
\begin{equation}
K = \sum_{i=1}^n K(z_i, \bar z_i)
\end{equation}
and that the local low energy effective field theory of massless modes when D-branes coincide is given by ${\cal N} =4$ SYM theory. This is the approximation in which the D-branes do not cause back-reaction in the background metric, and that when D-branes coincide locally, to leading order they feel as if they are in flat space.

\item
 If we separate a pair of D-branes infinitesimally, the W-bosons and massive off-diagonal short open string modes between them should have a mass that is proportional to the distance between the branes.
\begin{equation}
m^2_{ij} \sim d^2(i,j)
\end{equation}
where the mass of these modes is controlled by the string tension. Indeed, all polarizations of these particles should acquire this same mass. 
Requiring that this last property is true for arbitrary separations between the branes is probably too much to ask for (this was studied in \cite{DKO,Dcurved,DO} finding various problems with implementing this idea satisfactorily). 

\item
The fact that we can put all branes on top of each other suggests that we have a $U(N)$ invariant non-linear sigma model description of the full physics.

\item
When we consider the Calabi-Yau singularity, we expect that placing a D-brane there corresponds to regions of enhanced gauge symmetry, characterized by brane fractionation and to a quiver gauge theory with some additional properties. The most important such property is that the action is a single trace function. This last property comes from requiring that
the action of open strings is generated by disk amplitudes in string theory.
\end{enumerate}

Now, let us start with a given field theory, with such a single trace action, and we want to know how far can one go towards satisfying the above axioms given this information and the fact that the theory is superconformal. The structure required for the theory in order to match a low energy effective description of D-branes is that we have a supersymmetric quiver theory (the details can be found in \cite{Brev}). We will work only with oriented strings. This simplifies
the structure considerably, and the quiver nodes represent $U(n_i)$ gauge groups, while the arrows are bifundamental fields. The following arguments are very technical and can be skipped
if one assumes that the above structure holds. Our purpose is to show that these are consequences of superconformal invariance.

Now we can ask if we can prove that axiom one holds for this quiver theory. For axiom one, this has been addressed in the papers \cite{BLdt, BJL, BL, Brev, Bcon}. It was
noted there that solving the F-terms associated to single trace superpotential was
equivalent to solving for the representation theory of an associated non-commutative algebra
$\CQA$ derived from the superpotential of the theory. The letter ${\cal Q}$ is in the name as a mnemonic for the fact that we started with a quiver and a superpotential. Thus, one reduces the problem to 
calculating all the irreducible representations of this quiver algebra.

The typical representation is of the form
\begin{equation}
R = \oplus_i R_i
\end{equation}
From this information, one can in principle read the rank of the gauge groups and the classical vevs of the chiral fields.

Representations of the algebra that differ by a change of basis are equivalent. This is 
a remnant of gauge invariance. Thus, in the above direct sum the order of the summands does  not matter. This gives us a formal structure of the moduli space of vacua as being given by a direct sum of $N$ unordered irreducible representations. 

If the equivalence classes of irreducible representations come in continuous families forming a (complex) manifold $V$, then the
moduli space will be of the form
\begin{equation}
{\cal M}_N \sim \hbox{Sym}^N V
\end{equation}

This structure depends on having no extra branches of moduli space where the dimensions of the representations change. This phenomenon happens at singularities in $V$ where branes might fractionate. A set of conditions on the quiver algebra that seem to ensure that the representation theory does not have pathologies were written down in \cite{Brev}. One can check that the examples of orbifolds of flat space and the field theory associated to the conifold  satisfy all of these properties.

It is conjectured that if the quiver algebra derived from the superpotential is regular and finitely generated over the center,
and with some extra technical properties, then the space $V$ will turn out to be a Calabi-Yau three-fold. 

Given this information, we find that the moduli space is indeed described by $N$ particles in a Calabi-Yau manifold. Also, the structure of the representations give us a way to compute the unbroken gauge group. This uses natural structures on the space of representations. The idea
is that homological algebraic methods describe the effective physics (a generalization of the geometric homological statement by Douglas \cite{Dhom}). 

If $R$ is such a direct sum representation, the unbroken gauge group is 
\begin{equation}
Hom( R, R)_{QA}\sim \prod_j GL(n_j, \BC)
\end{equation}
where $n_j$ are the multiplicities of each irreducible representation. This is exactly as expected for $n_j$ coincident D-branes at a point.

We see this way that the structures required to match the D-brane picture arise quite naturally from the field theory itself, without invoking string theory for the construction. This is what we require if we expect to show that the AdS/CFT correspondence holds: the field theory alone should be enough to describe ll of the stringy dynamics in $AdS$ without resorting to 
string theory arguments. Thus, axioms one and two are easy to handle.

For axiom three we need more work. Of course, classically, once we have decided that
we look at spaces of direct sums of representations, making a choice of basis is a gauge choice. If the basis is diagonal, so that (schematically)
\begin{equation}
R \sim \begin{pmatrix} R_1 &0 &\dots\\
0& R_2 & \ddots\\
\vdots& \ddots & \ddots
\end{pmatrix}
\end{equation}
this is, all of the fields in the quiver are block diagonal by blocks associated to the representations, then we can show that the classical Kahler potential has the required sum 
form. This special choice of basis is a type of eigenvalue basis, where each $R$ represents
a block eigenvalue. The formal structure of the representation theory guarantees that this choice is possible so long as the irreducibles are all distinct. For non-generic cases one can 
have representations that are decomposable but not direct sums.

This is, they fit into short exact sequences
\begin{equation}
0\to R_1 \to R\to R_2\to 0
\end{equation}
where $R\neq R_1\oplus R_2$. These representations are off-diagonal as follows \footnote{In commutative algebra they represent some blowup of the singularity: the difference between a symmetric product and a Hilbert scheme}
\begin{equation}
R \sim \begin{pmatrix}R_1& X\\
0 & R_2\end{pmatrix}
\end{equation}
In the field theory, the $X$ represents some chiral field vevs connecting the diagonal blocks.
For each $R_1$, $R_2$, there is a diagonal $U(1)$ vector field under which the $X$ are charged, but not any field associated to the $R_1$ or $R_2$ block. A vev of $X$ will require a non-zero D-term for these $U(1)$ fields. This is impossible in the case of conformal field theories. A D-term would induce a mass scale. Thus, although these situations can happen holomorphically, they do not solve the D-term field equations and do not correspond to a true vacuum \footnote{These types of configurations do matter in the more general case of fractional branes, and they lead to Seiberg dualities \cite{BDoug} by changing basis of fractional branes via these constructions}. 

The second part of axiom three requires that the massless degrees of freedom when D-brane coincide are those of ${\cal N}=4 $ SYM for gauge group $U(2)$. This requires that
at the limit where the D-brane coincide we have three off-diagonal chiral fields that are becoming massless, plus the gauge field. The spectrum of massless chiral fields that are "stretching between branes one and brane two" can be computed using
homological methods as well
\begin{equation}
n_{ch, 12} \sim \lim_{R_1\to R_2} \dim_{
\BC}\hbox{Ext}^1( R_1, R_2) = 3
\end{equation}
This result follows if the quiver algebra is locally equivalent to $V$ as an algebra, in the sense of Morita equivalence. The numbers here indicate the coherent sheaf intersection numbers of a point in three dimensions (for more details about how points behave homologically see \cite{Asp}). 

The construction implies that near the degeneration locus where two of the D-branes coincide we have the massless spectrum of fields of ${\cal N}=4 $ SYM. We also need that the interactions are correct. Gauge invariance guarantees that the massless vector multiplet will couple to the other 
degrees in the usual manner (following an effective Yang Mills action for the enhanced gauge symmetry), with some effective coupling constant $g_{eff}$ for the Yang Mills interactions.
At this level this is a parameter of low energy effective field theory. In general one expects this $U(2)$ dynamics to be given by some diagonal embedding in the gauge group of a quiver diagram. Thus, in principle one can compute $g_{eff}$ from some other data as follows
\begin{equation}
\frac 1{g_{eff}^2}= \sum \frac{n_i}{g_i^2} \label{eq:geff}
\end{equation}
where the$g_i$ are the gauge coupling constants associated to the nodes of the quiver, and the $n_i$ are the ranks of the corresponding gauge groups $\prod U(n_i)$. The embedding of the global $U(1)$ that determines the unbroken group of a single brane is diagonal, this is what determined $g_{eff}$. For our purposes, only the value of $g_{eff}$ will matter

Since we are not near the tip of the cone of $V$, we expect that we can do an expansion in 
massless fluctuations $\delta\phi$ around a background field where the representations coincide.

However, we need to worry about the effective superpotential for the massless degrees of freedom. We can argue easily that to leading order we have the same superpotential as that of ${\cal N}=4$ in fluctuations. The reason for this is that there can be no mass terms in the superpotential (all particles are massless when the representations coincide). 
Thus the potential is cubic and higher order. If the effective superpotential for the massless degrees of freedom is not of the form 
\begin{equation}
\int d^2\theta \tr (\delta\phi_1[\delta\phi_2, \delta\phi_3])
\end{equation}
then diagonal matrices do not solve the F-terms generically, a contradiction with the shape of the moduli space that we already computed. This is a self-consistency requirement of the low energy physics (this argument can be found in a slightly different form in \cite{DKO}). 

Up to here, $\lambda$ is independent of $g_{eff}$. To get the correct dynamics of ${\cal N}=4 $ SYM, $\lambda$ and $g_{eff}$ must be related to each other. Such a relation follows from requiring conformal invariance of the low energy effective action near this point in moduli space \cite{LS}. We notice that if we change $g_{eff}$ and keep $\lambda$ fixed, the potential for the fields has $U(3)$ symmetry, but not $SO(6)$ symmetry. If we move in moduli space
to separate the two branes slightly the $U(3)$ global symmetry is broken down to $U(2)$.
Under this symmetry the massive vector field will be a singlet of $U(2)$, and there will also
be a doublet of massive chiral superfields charged under $U(2)$. The mass of the scalars is proportional to $\lambda$, while the mass of the $W$ is proportional to $g_{YM}$. If these numbers are not correlated, we have different masses for these modes.

Also, if $\lambda$ and $g_{YM}$ are not related appropriately, the one loop effective action 
will give rise to logarithmic divergences that correspond to wave function renormalization of 
the $\delta\phi$. Such logarithmic divergences should be absent in a Conformal field theory, even if the conformal group is spontaneously broken in the vacuum. Their absence should be a consequence of the conformal Ward identities \footnote{The argument as stated is reasonable and compelling, but it does not  constitute to a proof of the corresponding relation. Such a detailed proof has eluded the author in the general case.} 

Notice that now that we are using the full Conformal invariance of the  theory, we have found that the masses of the off-diagonal modes are degenerate in order to have local conformal invariance (we have a finite theory: to one loop there is no divergent wave function renormalization). Moreover, in ${\cal N}=4$ SYM the mass of the off-diagonal modes is proportional to the distance in moduli space separating the two 'eigenvalues'. The argument we have followed
has landed us automatically in the axiom four of D-geometry to leading order.

We also have to worry about corrections to the Kahler potential on the moduli space of vacua
when we integrate in and out these massless modes. However, in ${\cal N}=4$ 
SYM there are no corrections to the Kahler potential on the moduli space.  Thus any correction will have to come from higher order terms in the effective action. These higher order terms will be suppressed by another mass scale: we can call it either the scale of the radius of curvature on  the moduli space, or the scale of global conformal symmetry breaking. They are usually associated to integrating out the other modes of the field theory that are massive, even when the representations coincide: in the case of orbifolds, they result from integrating out fields connecting a brane with its images. They also result from non-linearities in the global Kahler potential for curved moduli spaces.

If this scale is very large compared to the typical masses of the degrees of freedom that were integrated out, there is no correction to the local effective Kahler potential. Unfortunately, 
this is not the whole story: in quantum field theory irrelevant operators can have large contributions and be dangerously irrelevant because of the possibility of quadratic divergences, etc. These can be there even for a single 'brane'. Conformal invariance can be invoked again here: the absence of quadratic divergences (or any divergences in the effective action on moduli space) should be a consequence of the Conformal Ward identities. 

The computation of quadratic divergences in $d=4$, under dimensional regularization are equivalent to logarithmic divergences of the dimensionally reduced theory in $d=2$. One can relate these divergences in $d=4$ to the computation of $\beta$ function for a non-linear $\sigma$-model with $(2,2)$ supersymmetry in $d=2$ associated to the same 
Kahler manifold. The vanishing of the $\beta$ functional in $d=2$ implies that the metric associated to $K$ is Ricci flat. This argument suggests that the metric associated to $K$ is actually a 
Calabi-yau metric. Given this information, the agreement of the masses to the metric distance
persists to the next order in the expansion (this was first observed in \cite{DKO}).

Axioms 5 and 6 seem to be built naturally in the formalism of the quiver theories: they are used as inputs to determine the structure of the theories, and once the theory has been constructed, it naturally has the required properties.

We have thus found that the metric on the symmetric product moduli space should be Ricci flat. Since the space is
Kahler, we are getting a Calabi-Yau metric. Moreover, we have that the dynamics of degenerations is locally that of ${\cal N}=4 $ SYM. This guarantees that there are no corrections to the effective Kahler potential on moduli space near the points of enhanced symmetry, at least to leading order.

\section{Self-consistent strong coupling dynamics}\label{sec:SC}

So far all of our discussion of the field theory has been classical, and we have argued that certain quantum corrections on the moduli space effective dynamics are absent because
of superconformal invariance of the theory. This constraints the metric of the quantum field theory on moduli space to produce a moduli space whose metric is induced from the Calabi-Yau metric on $V$.

We now want to go back to the field theory compactified on $S^3\times \BR$ and analyze it
at strong coupling from first principles.

We need to be careful with what we mean by strong coupling here. For most non-trivial conformal field theories, there is no perturbative classical vacuum at the origin of field space (the unique classical superconformal point in the moduli space of vacua). The 
reason for this is that the fields do not have canonical dimension and the Kahler potential is not quadratic at the origin of field space. Thus, in the sense that these are not free field theories one can argue that one is always at strong coupling.

What we want to do is analyze the theory on the moduli space of vacua away from the singularities of $V$. At these locations in field space, the low energy effective field theory is essentially ${\cal N}=4$ SYM, and the other degrees of freedom are massive. Thus one can expect to be able to use perturbation theory for the massive modes 
if one knows in detail the kahler potential and one can integrate them out systematically.

The question of strong coupling is then one of the effective ${\cal N}=4 $ SYM dynamics. 
We have an effective coupling constant $g_{eff}$, and $U(N)$ group, and we can choose to put all the branes on top of each other. Thus one can ask if the corresponding effective t' Hooft coupling is large or small. This is, if $g_{eff}^2 N\sim \lambda_{eff}$ is bigger or smaller than one. We want to analyze the theories in the regime where $\lambda$ is very large, but $g_{eff}$ is very small. This is the limit in which the dual gravitational theory is expected to have large radius (small curvatures) and where classical supergravity should apply \cite{Malda}.

As discussed previously, the chiral ring is special because it is guaranteed to be dual to supergravity solutions in suitable semiclassical limits. Thus, if we want to understand these solutions in field theory, we have already showed that all we need to do is look at the classical moduli space of vacua and analyze the relevant configurations.

As we have discussed, these configurations will be characterized by $N$ points on $V$.
If we think in terms of quantum mechanics, we will have a wave function of $N$ particles on $V$, let us call it $\psi$. Because the exchange of particles is a gauge symmetry (the ordering of points of $V$ does not matter in the moduli space), the wave function of any element of the chiral ring should be that of $N$ bosons on 
$V$. 

We should obviously also have a particular wave function $\psi_0$ for the ground state. We want to determine that effective wave function. It's properties will ultimately determine if the approximations that we make are self-consistent or not. 

Since the chiral ring is the holomorphic coordinate ring of the moduli space of ${\cal M}\sim \hbox{Sym}^NV$, there is a natural wave function on ${\cal M}$ to describe these states:
an element of the chiral ring $\CO_\phi$ is associated to a holomorphic function $f_{\CO}$ on ${\cal M}$. However, these functions are not $L^2$-normalizable on the moduli space ${\cal M}$. We should have that the wave function describing the state in the associated quantum mechanics is
\begin{equation}
\psi_{\CO} = \psi_0 f_{\CO}
\end{equation}
and therefore the ground state wave function basically determines the full structure of the chiral ring and how it relates to particular classical configurations.

This prescription is natural in the mathematical sense: we use an obvious multiplication
of functions, 
 and it is the natural generalization of the prescription that was given in ${\cal N}=4$ SYM in \cite{Droplet}.  A balance between  $\psi_0$ and $f_{\CO}$ will encode all of the important information about
how the wave function localizes around some particular point in moduli space. In this
sense, $\psi_0$ is crucial for an understanding of the detailed dynamics
of this approximation (e.g. the dictionary between elements of the chiral ring and precise classical configurations that they are dual to under the operator state-correspondence).
A precise description of $\psi_0$ is beyond the scope of the present paper, and will be taken 
up elsewhere, in the second installment of this series of papers \cite{BHart}. 
For the purposes of this paper all we need are some qualitative features of $\psi_0$, and 
analogies to the case of ${\cal N}=4 $ SYM, where such a prescription has been given and analyzed \cite{Droplet, BCV} (see also \cite{BCorr,BCott} for the cases of orbifolds).
The point of view that we take here is that the dominant effects don't depend on the details of $\psi_0$, but on the fact that we have a gauge theory and that the dynamics on the moduli space is essentially (locally on moduli space) the same dynamics as that for ${\cal N}=4$
SYM.

This type of description is good to describe semiclassically the set of states that saturate the BPS bound: we are just quantizing the relevant set of classical configurations. We also know that around general configurations of points in $V$, the masses of the off-diagonal modes are proportional to distances in moduli space between the particles. The constant of proportionality
is essentially $g_{eff}$. Thus, if on general grounds $g_{eff} d(u,v)$ are much larger than the natural 
scale of the $S^3$, namely $1/R$, where $R$ is the radius of the sphere, then these massive modes can be integrated out
systematically and the low energy dynamics (the modes whose energies are of order $1/R$) are described by effective field theory around configurations on the moduli space of vacua.
Let us name this type of configuration a {\em  massive off-diagonal}  configuration (MOD).

We want to argue that that the effective dynamics of gravity is closely related to MOD configurations, and that integrating out the off-diagonal modes systematically is a good approximation even for the quantum vacuum of the theory. Notice that classically the vacuum occurs at the origin of moduli space, where all ``D-branes are at the origin" and generically all distances $d(u,v)$ are zero, so we want to show that the quantum vacuum is far from the classical vacuum in a way that we can control. We also want to show that we are in a situation where perturbation theory around the correct set of configurations can be done systematically, at least for some set of degrees of freedom. We will call this the MOD principle hypothesis.

{\em MOD principle hypothesis:} for the vacuum state of the SCFT the effective low energy dynamics localizes to effective field theory around some typical configurations in the moduli space of vacua that are far away from the origin. Moreover these typical configurations 
have the MOD property for most of the off-diagonal degrees of freedom (the number of such degrees of freedom scales like
$N^2$ ). For these off-diagonal modes $g_{eff} d(u,v) >>1$.

The precise meaning of this statement is that if $g_{eff}$ is fixed and small, and we take $N\to \infty$, then the MOD property is true for all off-diagonal modes in the vacuum state. At finite $N$, or finite but large  $\lambda_{eff}$ for that matter, the fraction of off-diagonal modes that are not sufficiently massive
is controlled by some inverse power of $\lambda$. In the gravity setup, these conditions on $N$ and the coupling constant $g_{eff}$ tell us that
the radius of the geometry starts becoming comparable to the string scale and that some string modes are becoming light (their energy becomes comparable to the energy of gravitational quanta).

We want to argue that this hypothesis is self-consistent: that is, if we start with the MOD principle and find the relevant field configurations on moduli space that describe the vacuum (we solve the relevant moduli space effective quantum dynamics), then 
these configurations will satisfy the MOD property. Furthermore, we want to show that this can be used systematically to make some calculations of various non-BPS quantities in the field theory from first principles.

Since these quantum vacuum configurations are necessarily far from the classical vacuum, we can argue that we are analyzing the theory in a non-perturbative way, and that we are expanding around a non-perturbatively improved classical set of configurations.

Considering the fact that without something similar to the MOD principle (some setup where we are able to use perturbation theory around a reasonable configuration) we have very little chance of performing any calculation at all directly in a superconformal field theory, especially those that do not have a free field limit (e.g., the conifold), the MOD principle is placing us in a 
situation where we have a chance to perform calculations at strong coupling directly in field theory. From the point of view of the AdS/CFT, if we show that this description has a reasonable geometric interpretation, we might be able to understand the emergence of gravitational descriptions of a-priori non-geometric quantum systems. In essence, these type of arguments could pave a way to provide some sort of proof of the AdS/CFT correspondence for some general class of backgrounds.

Let us now begin the analysis of the relevant classical field configurations of the field theory on $S^3$, with the assumptions that we have made so far.

As discussed previously, we have argued that for the chiral ring only configurations of trivial gauge fields are relevant. Thus, we will set the connection 
to zero (this is a particular gauge choice).
We have also argued that
for these configurations all spatial gradients vanish. Thus the configurations that we need 
correspond to a dimensional reduction of the quantum field theory on the sphere to a point: 
only the rotationally invariant configurations are kept.

If we start from the full gauge theory and consider only these types of configuration, we have
that the dynamical system is some matrix quantum-mechanical model. We have not yet imposed fully the conditions for being on the moduli space of vacua: these require that
the D-terms and F-terms vanish. We will consider only constrained field configurations that satisfy this property. In principle, given the classical lagrangian, we can truncate classically to this set of configurations and we can calculate the induced classical dynamics exactly.

This is a subset of the possible configurations in the matrix quantum mechanics.
We will also keep the constant modes of the time components of the gauge fields $A_0$ to impose the gauge invariance on the set of classical configurations. Since these components are not dynamical (they just implement constraints), we are not adding degrees of freedom.

Given the classical lagrangian of the theory, we obtain this way some type of dynamical system on a constrained matrix model. Up to gauge equivalence, any such field configuration can be diagonalized to a description of $N$ particles on $V$ (remember in our notation $V$ is the classical moduli space of vacua). This can be understood by doing a classical 
diagonal ansatz. The ''eigenvalue" blocks become the only degrees of freedom that matter.

The classical dynamics on ${\cal M}$ is thus a non-linear sigma model for $N$ particles on $V$ in the presence of some classical potential: the Kahler potential itself (this is understood from \ref{eq:clham} by setting the corresponding terms to zero). Given the fact that in the classical Kahler potential the particles decouple (the Kahler potential is additive $K=\sum K_i$), and the sigma model depends only on $K$, as well as the potential (remember equation \ref{eq:confcoup}). Thus, classically, we have that the dynamics is governed by the following sigma model
\begin{equation}
{\cal L} \sim Vol(S^3) \sum_p  \int dt \left(g_{i\bar j}(z^p, \bar z_p) \dot z_p^i \dot {\bar z}_p^{\bar j} - K(z, z_p)   \right) 
\end{equation}
the other terms in the Lagrangian, that come from spatial gradients, D-terms, F-terms and gauge fields are automatically zero. Also we have introduced coordinates $z, \bar z$. These are suitable local complex coordinates for a particle near its position. In term of real variables we would use $g_{ab} \dot x^a \dot x^b$.

There is a natural real coordinate system on a Calabi-yau cone space where $g \sim dr^2 + r^2 \hat g_{ij} (X)$.  Here $X$ is the five dimensional real base of the Calabi-yau cone (the Einstein-Sasaki manifold), and in this coordinate system the scaling transformation just acts by rescalings of $r$. Thus it must be the case that in these coordinates $K\sim r^2 f(X)$, for some function $f$ on $X$. 
This has been studied in the paper \cite{MSY}, where they found that $f$ is independent of the coordinates on $X$. 
So up to a normalization factor we have that $K=r^2/2$.

In a naive quantization of the system all the particles are decoupled, so we can separate variables on the different particles. The one particle effective Schr\"odinger operator would be of the form
\begin{equation}
H_{eff} \sim -\frac 1 {2 r^5} \partial_r r^5 \partial_r - \frac 1 {2 r^2} \nabla_X^2+ r^2/2  \label{eq:Heff}
\end{equation}
We can furthermore separate variables between $r$ and the Sasaki -Einstein
variables.  In the above expression the powers of $r^5$ come from computing the radial dependence of $\sqrt g$ in these coordinates.

The lowest energy solution  of the one-particle problem will have a vanishing gradient on $X$, and 
we end up with the same Schrodinger problem of a particle in a harmonic oscillator in $6$ dimensions, in a situation where the angular momentum vanishes (only radial motion).
Thus, the problem ends up being identical to a single particle in a 6-d harmonic oscillator. This 
is the problem that one has to solve in ${\cal N}=4 $ SYM.

The ground state wave function will therefore decay as 
\begin{equation}
\psi_0 \sim \exp( - r^2/2)
\end{equation}
and if we consider all particles, then we have
\begin{equation}
\psi_0\sim \exp(-  \sum_p r_p^2/2)= \exp(-\sum_p K_p)
\end{equation}
we see that the wave function depends only on the Kahler potential in this preferred coordinate system. 

A more precise quantization would remember that we started from  a matrix model, and that
choosing diagonal variables involves a choice of gauge. Thus the full laplacian gets modified by measure effects: we need to compute the volume of the gauge orbit. Doing this in general seems very hard (although it has been computed for the case of ${\cal N}= 4 SYM$ and 
some orbifolds \cite{Droplet, BCorr,BCott}). 

However, we know that there are regions where the volume measure of the gauge orbit 
vanishes because we have enhanced symmetry (therefore the corresponding gauge orbit has lower 
dimension as it is invariant under extra symmetries). This occurs when two particles in $V$ coincide, or when one of the particles moves to the tip of the cone of $V$. The net effect of
this measure factor is to give an effective repulsion between the particles in the full wave 
function, and they are also repelled from the tip of $V$. This is a higher dimensional 
generalization of the repulsion of eigenvalues in matrix quantum mechanics due to the VanderMonde determinant measure \cite{BIPZ}.

Furthermore, one can argue that the full measure of the gauge orbit should be a function with some definite global scaling. This is because the set of configurations that satisfy the 
conditions to be on the moduli space is covariant under rescalings (all the conditions that are satisfied are homogeneous). Finally, one expects that the measure near the degeneration locus where two particles are near each other behaves just like it does in ${\cal N}=4 $ SYM, roughly, that it degenerates like $d(u,v)^2$. If we put all of points on top of one another (still at finite distance from the tip of $V$), then we would have a measure factor (we will call it $\mu^2$) which is roughly of the form
\begin{equation}
\mu^2 \sim \prod_{i<j} d^2(u_i,u_j)\label{eq:measure}
\end{equation}

This problem has been fully solved for ${\cal N}=4 $ SYM , where the equation (\ref{eq:measure}) is exact. Moreover one can find a full solution of the Schrodinger equation that includes the measure term, such a ground state wave function was called $\psi_0$ and it was just the Gaussian wave function: namely, the solution to the full quantum problem and the simplified problem without the measure were the same. However, even though the wave function looks the same, all vevs are different because the measure appears in the definition of probabilities, by taking
\begin{equation}
\vev{\CO} \sim \int \mu^2 dx |\psi_0(x)|^2\CO(x)
\end{equation}
where $x$ are the coordinates on moduli space. 

If we modify the wave function to absorb a square root of the measure, namely, we choose \begin{equation}
\hat\psi_0 \sim\psi_0\mu
\end{equation}
then we have the obvious equality
\begin{equation}
\vev{\CO}\sim \int dx |\hat\psi_0|^2 \CO(x)
\end{equation} 
The important thing is that for measurements, with the use of $\hat\psi$ the measure
of the gauge orbit does not appear anymore, and instead the measure is the standard one for $N$ particles on $\BC^3$: we have to do $N$ dimensional integrals, where the $N$ variables are factorized.

If we interpret the probability density associated to $|\hat\psi_0|^2$ as a probability density
for a statistical mechanical system, then we have a gas of $N$ particles moving on 
$\BC^3$. The corresponding partition function is
\begin{equation}
Z \sim \int \exp(-\beta \tilde H)
\end{equation}
and in this case, if we choose $\beta=1$ for convenience, then
\begin{equation}
\tilde H = \sum_p \vec x_p^2- \sum_{i<j} \log|\vec x_i-\vec x_j|^2
\end{equation}
where we have used the explicit formula for the distance in flat coordinates 
$d(u_i, u_j) = |\vec x_i-\vec x_j|$. This is, we have a gas of particles with some external confining potential $\vec x_p^2$, and repulsive logarithmic interactions in six dimensions.

For the case at hand, where the moduli space is described by $N$ particles on $V$, we will make the assumption that the same type of ansatz will work. Using the observations in \cite{Droplet}, it was shown that the gaussian was a solution of the Schr\"odinger problem because the measure was a homogeneous function. This tells us that the simple problem we solved above is the  complete solution to the ground state wave function in this moduli space approximation, so long as the measure term is a scaling function.
The idea is that the essential dynamics for the relevant configurations is in the case of $V$ is the same as ${\cal N}=4$ SYM. After all, locally on the moduli space of vacua the effective dynamics is that of the moduli space of vacua of ${\cal N}=4$ SYM, except for the tip of the cone $V$ (a set of measure zero). Globally, of course, the topologies are different.

Thus, given a solution of the trivial  problem (\ref{eq:Heff}), the full (measure modified) wave function will be
\begin{equation}
\hat\psi_0 \sim \psi_0 \mu
\end{equation} 
Given $\hat \psi_0$, the probability density of finding the particles at some locus is
\begin{equation}
p \sim |\hat\psi_0|^2
\end{equation}
and that the measure is now over $N$ copies of $V$, the same type of interpretation is possible. We expect a gas of $N$ particles on $V$, with some confining potential determined by the one body problem on $V$, with some effective logarithmic repulsion of the particles and also repelled from the tip of $V$.

The single trace property of the action, combined with planarity, suggests that the measure factorizes into a product of two-body problems. This has been explicitly evaluated in the case of some orbifolds \cite{BCorr, BCott}. As such, the measure is a product of two point measures.
The scaling property of the measure guarantees that the logarithm of the measure also scales as $\log(d(u,v))$ in the asymptotic regime. This is, the logarithmic repulsion 
controls the effective two body potential both at very short distances and at very long distances.

Now we have a calculation that we can do. Since the particles are confined  (by the $r^2$ term in the Kahler potential) and they have repulsive interactions, they will find some equilibrium configuration. We also expect that this equilibrium configuration will have some thermodynamic limit ($N\to \infty$), after proper rescaling of the variables,  where we
can calculate the density of particles, and the typical distances between them.

 We can estimate the 
size of this configuration by assuming all particles are at a typical distance of order $R$
from the tip of $V$, which depends on $N$. 

The confining potential will scale like $N R^2$, while the repulsive logarithmic potential will scale like $N^2\log(R)$. The balance of repulsion and confinement tells us that
\begin{equation}
\partial_R (NR^2-N^2 \log(R))=0
\end{equation}
or equivalently, that $R$ scales as $\sqrt N$. Notice that in this formula for $R$, $g_{eff}$ does not show up: when we write the classical effective field theory for the scalars 
on the gauge orbits associated to the moduli space the constraints that we impose remove
any dependence on $g_{eff}$.

Since the particles repel each other, we expect that the typical distance between a randomly selected pair of them will also scale like $R$, namely , for most pairs of particles we have that $d(u,v)\sim R\sim \sqrt N$.

If we put this information back to verify the self consistency of the MOD hypothesis, we find that the off-diagonal modes have a mass of order
\begin{equation}
m(u,v) \sim g_{eff} \sqrt N \sim \sqrt {\lambda_{eff}} >>1
\end{equation}
as governed by the axioms of D-geometry, and the precise calibration of the off-diagonal masses in field theory units.
The last inequality is correct because we are assuming that we are at very strong t' Hooft coupling . This is the relevant regime for the supergravity limit to be valid in the dual theory \footnote{
Notice that in all our analysis we have not used supergravity to do a calculation, but just to 
select the appropriate regime in the quantum field theory where we should do our calculations}.

Of course, the argument above is not true for every pair of particles in the gas: there can be particles that are nearby each other in a gas, and for those, the typical distance will scale like some smaller power of $N$. The masses of the off-diagonal degrees of freedom associated to these nearby particles can be small.
 Let us say that we want to keep all modes that have 
 \begin{equation}
 g_{eff} d(u,v) <A /R 
 \end{equation}
where $R$ is the radius of curvature of the $S^3$, and $A$ is some scale. Let us also fix one particle $u$ from our gas. The number of those other particles that are relevant depends on $g_{eff}$ and $N$, since $d(u,v)$ for nearby points scales with some power of $N$.

For these nearby points, the dynamics is just like ${\cal N}=4$ SYM, with coupling $g_{eff}<<1$. The number of sufficiently nearby points is much much smaller than $N$, so we have an effective local 't Hooft coupling constant that can be much smaller than one, and perturbation theory for the light off-diagonal modes is valid. This is exactly what we wanted to show.

If we keep $g_{eff}$ fixed and tiny, and we take $N\to \infty$, all the off-diagonal masses
will go to infinity in the typical vacuum configuration. We have thus verified that the
vacuum is technically a MOD configuration and that the MOD hypothesis is self-consistent.
We have to keep in mind that the MOD hypothesis for the vacuum and for $g_{eff}$ fixed requires $N$ to be large, so we are naturally forced to consider a large $N$, strong coupling limit in the sense of t' Hooft \cite{'tH} in order to be able to make computations in this setup. 

Now, we want to understand better what the distribution of this gas of particles will look like and we would like to see what the simplest predictions of this formalism look like. This problem has been understood in the case of ${\cal N}=4$ SYM \cite{Droplet,BCott2}.

\section{Emergent geometry and BMN limits for the conifold}\label{sec:emg}

We have found so far that we have a model for describing the vacuum state of a conformal  field theory on $S^3$. The field theory is divided into two classes of fields: 
moduli directions and off-diagonal modes.

The off-diagonal modes are heavy and can be integrated out systematically. They are to be treated perturbatively. To leading 
order only the lightest modes could contribute to the effective action of the moduli. Because we
are working in supersymmetric field theories, we do not have to worry about vacuum energy corrections, nor about generating an effective potential on the moduli space. Non-renormalization theorems for supersymmetric vacua are expected to 
prevent these corrections.

 Integrating these off-diagonal modes out could generate corrections to the Kahler potential, but these are absent as the very light massive modes come in multiplets of ${\cal N}=4$ SYM. 
 This was suggested to be as a consequence of conformal symmetry. 
 Thus to leading order the moduli
dynamics decouples from the off-diagonal modes.

Because the theories under consideration are gauge theories, the moduli space is a set of gauge equivalence classes of configurations in the dynamical system we are considering. This is, 
\begin{equation}
{\cal M} \sim {\cal U}/G
\end{equation}
So we have the natural fibration
\begin{equation}
\begin{matrix} { \cal S} &\to&{ \cal U}\\
&&\downarrow  \\
&&{\cal M}
\end{matrix}
\end{equation}
where ${\cal S}= G// H$, where $H$ is the commutant of $G$ for the typical moduli space
point, namely $U(1)^N$.

When we quantize the system, we have a Schr\"odinger problem for some sigma model in ${\cal U}$, that has $G$ realized by isometries. Imposing the gauge constraints, the wave function becomes independent of the variables that parametrize the fiber ${\cal S}$ and it can be described by a wave function that depends only on the coordinates of ${\cal M}$. However, expectation values of observables
in the theory are naturally calculated in ${\cal U}$. To get observables over ${\cal M}$, we have to integrate over the fiber, and this produces a non-trivial measure in ${\cal M}$, that we have called $\mu^2$. The volume of the fiber degenerates at the special locus where the group $H$ displays enhanced symmetry: the dimension of the fiber changes at these points to a fibre of lower dimension.

Also, we have found ${\cal M}\sim \hbox{Sym}^N V$. We have argued by analogy with ${\cal N}=4$ SYM that the wave function takes the simple form
\begin{equation}
\psi_0 \sim \prod_{i=1}^N \psi^1(z^i, \bar z^i)
\end{equation}
where $\psi^1$ is a {\em one particle wave function}, that decays like a gaussian asymptotically. 
 We have also assumed that the measure factorizes
 \begin{equation}
 \mu^2 \sim \prod_{i<j} \mu^2(u_i, u_j)
 \end{equation}
 a fact that has been verified for ${\cal N}=4 $ SYM and it's orbifolds. This would also happen if the measure is calculated only via a one loop effect.

 There is also a possible measure associated to the fact that $V$ itself is a geometric quotient
$V\sim \tilde V/ G_1$, where $G_1$ is the gauge group of the ``field theory of a single D-brane". We have ignored this measure so far, except to state that it vanishes at the tip of the cone $V$, where we have enhanced gauge symmetry due to brane fractionation. We can always absorb this measure in the definition of $\psi^1$.
Since ${\cal M}$ is a symmetric product, there is a simple measure on ${\cal M}$: a factorized measured on each of the $V$. We chose to absorb the measure $\mu^2$ into $\psi^0$ by stating that
\begin{equation}
\hat \psi_0 = \psi_0\mu
\end{equation}

The square of the wave function on ${\cal M}$, $|\hat\psi_0|^2$ can  be represented as a probability distribution associated to a Boltzmann gas of particles on $V$ with two body interactions, characterized by the potential $\log(\mu(u_i, u_j)^2)$. 

We have also argued that in the limit where particles coincide, $\mu^2(u_i,u_j)$ vanish as
the square of the metric distance between them $d(u_i,u_j)^2$. When we take the logarithm of this degeneration, we have logarithmic repulsive interactions at short distances. We also argued that $\mu(u_i,u_j)$ should be a scaling function under dilatations, so that the logarithmic repulsion is also true at long distances. This is the same behavior that the associated gas of particles for
${\cal N}=4 $ SYM has. That corresponds to the special case $V= \BC^3$.

Locally, the gas of particles in $V$ 
behaves exactly as it does in the case of ${\cal N}=4 $ SYM, so we would expect that the fact that we replaced $\BC^3$ by $V$ is for the most part irrelevant, and that the gas settles down to a configuration that is similar to the one that one gets in the ${\cal N}=4$ SYM theory. Here we are proceeding by analogy.

In ${\cal N}=4$ SYM the gas of particles in the thermodynamic limit becomes a real codimension one surface on $\BC^3\sim \BR^6$. By symmetry of the gas partition function, one can argue that this surface is a sphere, where all particles are at the same distance from the origin $d= \sqrt {N/2}$ \cite{BCV}, and the density of particles in this surface is constant.

We want to argue that something similar happens on $V$: all the particles in the thermodynamic limit form a surface on $V$. The particles are repelled from the tip of $V$ at short distances, as well as being repelled from each other. The particles are also pushed at large distances towards the origin by the fact that the one particle wave-function in the Schr\"odinger problem is the exponential of the Kahler potential.

We find that none of the particles should be near the tip of $V$, but that all should be roughly at the same distance fro it. We would also expect that the particles would try to spread as much as possible without increasing their one particle energy by much, this is, we expect the configuration of this surface should surround the tip of $V$. This is, it has the topology of $X$.

Taking $N$ large shows that the typical distance of all particles from the origin is of order ${\sqrt N}$, so the field theory volume of this surface is $N^{5/2}$, and the surface area per particle is of order $N^{3/2}$.  However the details of this configuration depend on knowing the wave function in more detail. 

So far, we have a topology and a typical distance from the origin in field units. We also 
know that the off-diagonal mode energies are comparable to distances between particles.
If we excite a small number of these off-diagonal degrees of freedom perturbatively, the Gauge invariance  
under the unbroken $U(1)^N$ requires that the total charge vanishes. We can picture these off-diagonal degrees of freedom by a line joining the two particles that it connects. In this the off-diagonal degrees of freedom will form  closed polygons (just like in \cite{Droplet}).
The energy associated to
these degrees of freedom would be roughly the metric distance of the closed polygon (the length of the polygon). This is what we would expect for closed string states that follow the same polygon in $V$, or $X$ for that matter. These polygons, suitably dressed, where also found to describe certain strings in the dual geometry \cite{HM}.  

Thus the spectrum of states above the ground state has moduli excitations, and off-diagonal excitations. The moduli excitations need to be symmetric functions of the moduli variables. 
If these behave like polynomials (if we have similar behavior to that of  the chiral ring), then one can show that for low degrees of the polynomial a basis of traces of polynomials of a single variable give rise to an approximate Fock space description. In the chiral ring case, the degree is the dimension of the corresponding operator. This should be done in general.  (The notion of approximate Fock space was presented in more detail \cite{Bgg}. Those ideas might help define precisely what we mean by such a characterization of an approximate Fock space.).

The idea that traces can be treated as oscillators was used by Witten to describe the Fock-space of gravity in terms of protected operators in ${\cal N}= 4 $ SYM. This approximate orthogonality of multi-traces is expected from large $N$ factorization in matrix models. However, we have not proved these properties here yet. Numerical Monte-Carlo simulations in the special case of ${\cal N}=4 $ SYM show that the thermal fluctuations of the gas of 
particles for these chiral ring single trace modes are a collection of decoupled Gaussians for large values of $N$, so we would expect that this property is robust when we consider the case of particles in $V$ instead of particles on $\BC^3$. One would like to believe that these have to do with the propagation of "sound" in the gas of particles. From the point of view of geometry, this is exactly the place where one
expects to find gravitation. As we have argued before, in the case of the AdS geometry the chiral ring configurations have to be supergravity solutions, so this is a natural proposal.

\subsection{Locality and metricity}

We have argued that the gas of particles in the thermodynamic limit forms a surface inside $V$.
The topology of the shape of this surface is $X$. This surface has an induced metric from $V$. We measure this metric by turning on off-diagonal degrees of freedom, as the energies of these degrees of freedom are sensitive to the metric in a rather obvious way \cite{BCV}. Thus the metric property of the surface $X$ is important to describe physical excitations of the system.

If the surface $X$ fluctuates in shape, the spectrum of off-diagonal fluctuations will change, and 
is sensitive to the
induced metric on the new surface $X'$. This new metric will be reflected in how the polygon energies respond to it.

Now, calculations in ${\cal N}=4$ SYM have been done under the assumption that to leading order the particle density does not respond to excitations of the off-diagonal degrees of freedom. This should apply in the general case we are discussing as well. If we have added off-diagonal excitations in two different locations, the lack of backreaction of the density tell us that these off-diagonal excitations are excitations around a fixed background.
We want to estimate how these degrees of freedom interact. As we have argued, we
treat the off-diagonal modes perturbative.
Thus, with this assumption the  residual interactions between two excitations that are separated by some distance in $V$ should result from integrating out the 
other off-diagonal modes between the two locations, as the moduli are not contributing to this order. However,  
integrating out modes gives us energy denominators in time independent perturbation theory, and these denominators depend on the distance between the two off-diagonal excitations:
the farther away, the larger that the energy denominators become and we find that far away excitations are decoupled from each other.

This is, we find features that are compatible with an approximately local description of the physics. At 
intermediate distances (much smaller than the size of $X$, but much larger than the scale of granularity), the effective physics of the off-diagonal modes is indistinguishable from that of the case of ${\cal N}=4 $ SYM. The non-renormalization theorems of ${\cal N}=4$ SYM guarantee that the effective potential between excitations at intermediate distances decay by some power of the distance (getting the right power laws for the decay of the interactions with separation is the essence of Matrix theory proposal for recovering M-theory on the lightcone \cite{BFSS}. See also \cite{Wati} for how these depend on dimension). This is a hint that there are massless modes being exchanged between  the perturbations, because their interactions do not decay exponentially. Together with a
 combination of the next to leading effect from backreaction of the moduli, the effective 
 potential could in the end reproduce the full gravitational interaction between massive objects. 
 
 Unfortunately, such a calculation has not been performed, and it is beyond the scope of the present paper. However, we can speculate regarding the consequences of such a calculation.
 We should imagine that the corresponding calculation can be performed in the case of ${\cal N}=4 $ SYM. But if it can be done in this case, then for all the other cases the result of ${\cal N}=4 $ SYM is 
 a reasonable approximation in some regime.
 
 If we imagine that the result gives type IIB supergravity for the maximally supersymmetric case, 
 then the results that one gets with supergravity should have a diffeomorphism invariance symmetry
 associated to them. Such a symmetry would  persist
  in the other cases where we have ${\cal N}=1 $ supersymmetry, and we must conclude that we get the same supergravity theory as in the case of ${\cal N}=4 $ SYM. The difference between them will be at the level of global topology, not local dynamics.
 
 If there are corrections that need to be made, we can expect them to be related to the details of curvature in moduli space, because these enter the effective action of the off-diagonal modes in a direct manner. It is possible that 
 a systematic procedure to evaluate these corrections could be devised. These would be to leading order the $\alpha'$ corrections to gravitational interactions if we restrict ourselves to one loop order (or more precisely, planar diagrams)
in off-diagonal modes.

For this paper, we will consider a simpler problem that is tractable already. We will calculate the 
spectrum of strings in a plane wave limit, assuming that all the analogies that have been made with the special case of ${\cal N}=4 $ SYM are true. This calculation will be done at strong coupling, and we will
consider for simplicity the case of the conifold of Klebanov and Witten \cite{KW}. This is the simplest conformal field theory that one could consider that does not have a free field limit and where calculations in the field theory are essentially only known for the dimensions of operators belonging to the chiral ring and the conserved currents. 

\subsection{The conifold and BMN limits}\label{sec:con}

We are finally at a stage where we can discuss the implications of the formalism that has been set up before to the calculation of energies of strings in a special case of the dual field theory to $AdS_5\times T^{11}$.

We choose $T^{11}$ because the symmetry alone is enough to guarantee that the particle (eigenvalue) distribution in the thermodynamic limit is identical (isometric) to $T^{11}$. After all, $T^{11}$ is a homogeneous manifold: any point on $T^{11}$ can be rotated to any other point by isometries. Since the eigenvalue distribution on the conifold is a manifold of real 
codimension one and since the distribution has the same symmetries as $T^{11}$, the only choice is for the particles to form a surface that is isometric to $T^{11}$, and not just topologically equivalent. If we can guarantee the same for all geometries $AdS_5\times X$, 
that the eigenvalue distribution is equal to $X$ as a metric space embedded in the cone $V$, the the arguments below also apply (with some  minor modifications).

The idea is to set up a calculation where all we need to do is evaluate the energy of some short off-diagonal mode  $\beta^{i}_j$ \footnote{We call these modes off-diagonal, because we choose the
moduli space configurations to be in block diagonal form. The off-diagonal modes are essentially short strings stretching between  two D-branes (eigenvalues).}.
 However, the gauge symmetry of the wave function always requires that the full wave function is invariant under the unbroken $U(1)^N$, so if we add an off-diagonal mode $\beta^{i}_j$, we need to 'close the loop" and form a closed polygon. 

Moreover, we also have residual gauge invariance with respect to the permutations of the eigenvalues (particles in $V$), so a typical wave function has to be of the form
\begin{equation}
\sum_{ij} \prod \beta^{i}_j  \vac \sim \tr \prod \beta \vac
\end{equation}
We can improve how the eigenvalues are understood by adding functions of the coordinates
to the sum
\begin{equation}
\sum_{ij} \prod  f(u_i) \beta^{i}_j g(u_j)  \vac \sim \sum_{fg}\tr (f(Z) \beta g(Z)\dots)
\end{equation}
These wave functions of the particles (eigenvalues) are most easily understood of the chiral ring. Also, experience with ${\cal N}=4 $ SYM suggests that we should take a BMN limit where $f(Z) \sim Z^n$ for one complex coordinate on the moduli space \cite{BMN}. We will take $Z$to be diagonal.  We should consider states of the form
\begin{equation}
\sum_{n} \tr( Z^n \beta^\dagger Z^{J-n} \tilde \beta^\dagger \exp( 2\pi i k n/J))
\end{equation}
where we have two raising operators for off-diagonal oscillators $\beta^\dagger, \tilde \beta^\dagger$ in the low energy effective ${\cal N}=4$ SYM theory turned on.

 Notice that the above is gauge invariant under permutations of the eigenvalues, and that the $U(1)^N$ charges of $\beta$ and $\tilde \beta$ are opposite, since $Z$ is diagonal.

The trace function plays both the role of summing over permutations of eigenvalues and guaranteeing that the $U(1)^N$ charges cancel, as this is the only way to obtain a matrix with non-zero entries on the diagonal.

Notice that the description above is valid in the effective field theory limit, so long as the oscillators $\beta$ correspond to small distances on $X$. Now, we should pick a suitable (homogenoeus) holomorphic coordinate $Z$ on $V$.

We want to study the above wave function under the assumption of no-backreaction on the eigenvalue distribution. We find that under those conditions $Z_i^n  Z_j^{J-n}$ gives us the relative importance of the Ket in the off-diagonal degrees of freedom. The ones that
dominate maximize $|Z|$ for both oscillators \cite{BV}. The values of $|Z|$ is a function on $X$, so this maximization gives us a special locus on $X$.

 Ideally $|Z|$ on the surface $X$ attains a unique maximum, but the homogeneity on $u$ implies that $|Z|$ is the same for a full circle. After all, $X$ has a preferred $U(1)$ symmetry (the R-charge), and homogeneity of $Z$ implies that $u$ is an eigenvalue function under the $U(1)$ rotations. Thus, we find that $|Z|$ localizes the eigenvalue $i$ and $j$ to lie on a particular circle (a single orbit of the R-charge vector field).
 
 The large $J$ limit also implies that we can use a stationary phase approximation to find which pairs of eigenvalues dominate.  Thus, the relative phase between $z_i$ and $z_j$ is
 determined by $2\pi k/J$. Here we are copying the same arguments presented in \cite{BCV,BV}, with a modified notion of what $Z$ is. Since our choice is holomorphic, the 
 background $Z$ fields are in the ``chiral ring'' and must excite motion along R-charge orbits.

Thus, the relative position of $i,j$ are completely determined in the large $J$ limit (still $J$ should be sufficiently small relative to $N$ to avoid backreaction on the eigenvalues).

For the case of the conifold, the complex coordinates can be given as $uv= wz$. We want to expand around a simple configuration, so we choose the function $Z=u$. The maximum of $u$
occurs on $T^{11}$ in the locus where $w=z=0=v$. Indeed $T^{11}$ corresponds to a surface of the form \cite{CdelaOss}

\begin{equation}
|x_1|^2+ |x_2|^2 +|x_3|^2 + |x_4|^2= Const
\end{equation}
where $x_1 = u+v$, $x_2= u-v$, $x_3=w+z$, $x_4=w-z$,  and where the conifold variety is the hypersurface in $\BC^4$ determined by
\begin{equation}
\sum x_i^2=0
\end{equation}
Clearly $|x_1^2|+|x_2^2|$ is maximized when $x_3=x_4=0$, and then it is equal to $2 |u|^2$, as we must choose either $u=0$ or $v=0$ in order to solve the equations.
For other cases, where $v\neq 0$, we have that since $2u= (x_1+x_2)$, then 
\begin{equation}
4|u|^2 = |x_1|^2+ |x_2|^2 +2 |x_1||x_2| \cos(\theta_{12})< 2(|x_1|^2+|x_2|^2)
\end{equation}
where we use $2|x_1||x_2|< |x_1^2|+|x_2|^2$, a consequence of the arithmetic-geometric mean, and $\cos\theta\leq 1$.  This shows that in order to maximize $|u|$ the conditions described above are all necessary.

For us, $u= A_1B_1$ in the Klebanov-Witten  theory \cite{KW}, and it's conformal weight is $3/2$. Thus, the energy of the above is $3/2 J + E_\beta+ E_{\tilde \beta}$. We want to compute $E_\beta$. 

For this, we need the angle of separation  between $i$ and $j$ to be small, so that the approximation of ${\cal N}=4 $ SYM is valid, which is the limit in which we have argued 
that we know how to control the theory. This is most conveniently done when we take $J$ large,  but keep $n$ finite. Thus the distance between the two points on $T^{11}$ is $n/J$
times the radius of the circle fibre. As we have discussed, the radius of the fibre is of order $\sqrt N$, so the mass is of order $g_{eff}\sqrt N n/J $.

More precisely, there are two contributions to the mass of $\beta$. This is because we need to expand the Kahler potential to second order in $\beta$ also, because we have a coupling in the lagrangian of the form
\begin{equation}
- \int_{S^3} K
\end{equation}
from the conformal coupling of the metric to the Kahler potential.

We find that in expanding $K$ to second order, which is necessary both for the kinetic term and the potential, then the masses of the $\beta$ oscillators are exactly
\begin{equation}
 m_\beta^2 = 1+ a \lambda_{eff} n^2/J^2
\end{equation}
where $a$ is undetermined so far, because we do not know the exact size of $T^{11}$, only it's scaling with $N$. The factor of $1$ comes from the conformal coupling to the metric and 
it ends up being identical to the one for fields in ${\cal N}=4$ SYM theory. For fermions and gauge bosons there is no such coupling, but then they have the same effective masses.
This can be understood if we twist the Hamiltonian so that it is $H-J$. 
From the superconformal algebra $Q_\alpha$ and $S_\alpha$ anticommute to $H-J$, and the angular momentum act as a ``central charge''. Since the BPS wave functions do not affect $H-J$, one expects by the usual representation theory of this smaller subalgebra 
that the masses of bosons and fermions will be degenerate (up to the kinematics associated to
commutators of  $Q$ and $H-J$). Some of these differences can be absorbed into the $SU(2)\times SU(2)$ global symmetry quantum numbers, and one obtains results that look more similar to the plane wave of ${\cal N}=4$ SYM.

The functional form of this result coincides with the plane wave limit in the gravity theory 
\cite{IKM}, if we fix $a$ to the right value. This should follow from the exact wave function
and the measure and is analyzed elsewhere \cite{BHart}.
This is a very non-trivial result that we were able to derive purely from field theory considerations. The full spectrum of BMN excitations follows, as we have found that the result is the same as for ${\cal N}=4 $ SYM.

At this stage it is easy to consider other spherical harmonics of spin $s$ of the field $\beta$ on $S^3$, as in  \cite{BV} (these arise in the $(s/2,s/2)$ representations of $SU(2)\times SU(2)$ isometries of the sphere), giving us
\begin{equation}
m^2_{\beta_s}= (s+1)^2+ a \lambda_{eff} n^2/J^2
\end{equation}

This dependence on $s$ is the same as that suggested by integrability \cite{Dorey}. However, the string on 
$AdS_5\times X$ is not integrable, but the sigma model limit of a string in $AdS_5$ is. Since $s$ is related to spin in the AdS region, this suggests that the emergence of $AdS_5$ in all of these superconformal theories is universal, and should be essentially the same physics as the one found in ${\cal N}= 4 $ SYM. 

\section{Discussion}\label{sec:disc}

In this paper we have shown how to bootstrap the study of certain large $N$ conformal field theories at strong coupling, in particular those that are dual to $AdS_5\times X$ geometries, where $X$is a Sasaki-Einstein manifold. We have studied these field theories compactified on an $S^3\times \BR$ by using Hamiltonian methods.

We have argued that their strong coupling low energy dynamics is dominated by field theory configurations that explore the moduli space of vacua of the theory in flat space, where certain aspects of the theory can be treated perturbatively. We have shown that the relevant configurations are far away from the naive classical ground state.

This can be understood as a non-perturbative effect due to solving the theory with some quantum corrections exactly, in a similar vein to the spontaneous symmetry breaking induced by radiative corrections studied by Coleman and Weinberg \cite{CW}.
The main source for this non-perturbative effect is a measure factor 
that reflects that the moduli space of vacua arises from some gauge orbits which affect the quantum dynamics, even though classically we obtain a simple $\sigma$-model on the moduli space. This effect is a generalization of the effective repulsion of eigenvalues 
in matrix models \cite{BIPZ}.

 The classical moduli space under consideration is very closely related to $X$. In particular, $X$ is the base of a Calabi-Yau cone $V$, and the corresponding moduli space is given by a collection of $N$ particles on $V$. This is
 \begin{equation}
 {\cal M} = \hbox{Sym}^N V
 \end{equation}
 
 We have argued that the superconformal invariance of the field theory guarantees that the effective dynamics on moduli space satisfies the axioms of D-geometry \cite{Douglas}. In particular, we have shown that the associated D-particles are repelled from the tip of the cone in $V$ and that they form some configuration where all the particles are roughly at the same distance from the origin. This distance is of order $\sqrt N$ in field theory units. The dynamics that dominates is associated to the motion of the D-particles in moduli space, and their typical degenerations: where two D-particles coincide. This dynamics is locally equivalent in $V$ to the ${\cal N}=4 $ SYM theory. Thus all of these theories have essentially the same low energy effective dynamics. We have used this to explain why all these seemingly different conformal field theories are dual to solutions of type IIB string theory. If the ${\cal N}=4 $ SYM has this property, we have the same local dynamics in all of these models.

This is a particular form of background independence in string theory. We can phrase this point of view in the following terms: Background independence is tied to having locally on a distinguished slice of field space a universal structure for dynamics and interactions. The wave function of the universe on this distinguished slice and how it related to the rest of the degrees of freedom supply the non-trivial global aspects of the supergravity background: the global topology, metric, etc. In particular, given the supergravity bubbling solutions from \cite{LLM} and it relation to field theory \cite{Btoy}, it is clear that it is not just the ground state, but a large collections of chiral ring states in a fixed theory that share this property.

We have also shown that the universality of dynamics also translates into a calculable string spectrum in certain limits from field theory. In particular, we have shown that we can reproduce non-trivially the spectrum of open string on the  plane wave limit of the Klebanov-Witten theory \cite{KW}, matching the functional form of the supergravity limit calculations \cite{IKM}.
This can be done even in cases where the full string sigma model in the geometry is not integrable. Other approaches based on integrability \cite{BDS} would not work in these cases.

The formalism presented here has a lot of explaining power and permits one to do calculations in situations that were thought to be out of reach, like the case of conformal field theories where the fundamental fields have large anomalous dimensions and where there is no gaussian fixed point around which one can expand.

The description that we have given so far is, however, somewhat incomplete.

First of all, the measure factor required for some of the analysis was not given precisely, making some of the answers qualitative, in particular we did not show that we could recuperate the metric of $X$ precisely in string units. Doing this carefully requires new ingredients and will be taken up in the second paper of this series \cite{BHart}.

 Also, for conformal field theories the gauge groups are usually given by products of non-abelian factors $\prod SU(N_i)$, rather than $\prod(U(N_i))$, which is where the moduli space of vacua exactly matches a symmetric product \cite{Brev}. This is because the beta functions of the $U(1)$ factors make the $U(1)$ coupling constants run to zero and decouple in the infrared. This is also discussed in detail in \cite{IS}.
 Some of the $U91)$ fields are also massive due to 
the Green-Schwarz mechanism for anomaly cancellation. The extra polarization is
provided by a close string mode coming from the B-field around some cycle \cite{DSW} (this is also discussed in \cite{DM}). When we look at the 
IR fixed point physics all these massive and decoupled modes have to be removed from the field theory description.

The correct moduli space is in the end a $(\BC^*)^k$ fibration over the moduli space ${\cal M}$ that we have discussed above. Here $k$ is the number of nodes in the quiver diagram minus one (a global $U(1)$ is always decoupled). One can think of the problem this way because the $D$-terms for the $U(1)$ factors and the corresponding gauge constraint is absent. Thus the moduli space above is a symplectic quotient of the big moduli space

\begin{equation}
{\cal M}= {\cal M}_{big} // U(1)^{k}
\end{equation}
The fibers are parametrized by di-baryonic operators. These are not gauge invariant if
the $U(1)$ factors are gauged. 

Fortunately, most of the analysis of singularities in the base 
of ${\cal M}$ has essentially the same description in the big moduli space. The effect 
of these extra directions in moduli space can be thought of as having blown up the singularity
with a parameter that one can vary. 
As far as the axioms of D-geometry are concerned, we are modifying slightly the background where the D-branes are moving, but this does not affect their basic structure.
Moreover, we should notice that this effect is subleading in a $1/N$ expansion because it involves $O(1)$ degrees of freedom rather than $O(N^2)$ degrees of freedom. Thus, in keeping with the notion of using only leading contributions in powers of $N$, we can ignore these subtleties. Obviously these will be important for the description of the dual states to di-baryon operators. These are described in the dual gravity theory in terms of D-branes wrapping certain cycles \cite{GRW} ( see also \cite{BHK} for more details). This deserves further study.

Another issue that we have to understand is the notion of the radial direction in global $AdS$.
This direction is not immediately apparent in the study of the quantum field theories as we have done 
above. It is expected that this direction is related to the renormalization group of the field theory \cite{SuW} and therefore locality in this direction is hard to understand.
The radial direction can be explored with BPS states by looking at certain giant gravitons
\cite{HHI}. Their dual field description has been elucidated in \cite{CJR,Btoy}. 
These have to correspond to removing one D-brane from the gas of particles we have described and giving it a lot of energy. This will force the D-brane to explore the radial direction in moduli space. This indicates that there should be a close connection between the radial direction in moduli space and the renormalization group of the theory, but what that relation may be is completely mysterious at the moment. Without a proper understanding of how the degrees of freedom organize themselves radially it is very hard to understand from the field theory point of view why the problems of confinement and renormalization group flow can be described geometrically. 

More precisely, one could ask if it is possible to derive the string sigma model from field theory.
Some progress has been made in perturbative cases \cite{Kruc}

If we are willing to analyze only more general conformal field theories, and postpone the problem of having a field theory with a scale, one should consider setups 
where one turns on other supergravity background fields, like in the case of the 
Lunin-Maldacena solution of supergarvity \cite{LM}. These deformations where studied in the field theory in \cite{BJL, BJL2}, where it was shown that they destroy the moduli space of vacua of branes for general values of the deformations. Moreover it was shown that the field theory structure that governs these
setups is given by non-commutative geometry. At the level of perturbation theory,
these deformations modify the spin chain model of ${\cal N}=4 $ SYM \cite{MZ,BS} and they sometimes lead to a
string with twisted boundary conditions \cite{RBC} (see also \cite{Frol}). Some special cases 
where the moduli space of D-particles is not completely destroyed, this can be analyzed 
also at strong coupling \cite{BCorr} as done above.

If we consider that we have obtained a notion of locality in $X$ that is based on energetic considerations, we can also consider building initial state configurations that are localized in $X$ and might lead to interesting physics. In particular, it should be possible in general to setup initial condition for some small amount of matter localized in $X$ that might collapse in the dual geometry and produce a black hole ( a first attempt at such initial conditions has been presented in \cite{V}). This problem can then be analyzed in the field theory and we can ask what happens to geometry when we study this type of configuration. One should also be able to study heavy probes in $X$ separated by some distance and obtain the gravitational interactions between them. One would expect that this procedure might be calculable by integrating heavy (off-diagonal) degrees of freedom like in matrix theory \cite{BFSS}.

\section*{Acknowledgements}

I would like to thank R. Corrado, S. Vazquez and especially S. Hartnoll for various discussions related to this project. 
I would also like to thank the Weizmann institute of Science for their hospitality. It was during my stay there that various of the ideas that made this paper possible were hatched. 
Research supported in part by the DOE, under grant DE-FG02-91ER40618.

 \end{document}